# How to View an MCMC Simulation as a Permutation, with Applications to Parallel Simulation and Improved Importance Sampling

Radford M. Neal

Dept. of Statistics and Dept. of Computer Science
University of Toronto

http://www.cs.utoronto.ca/∼radford/
radford@stat.utoronto.ca

30 April 2012

**Abstract.** Consider a Markov chain defined on a finite state space, $\mathcal{X}$, that leaves invariant the uniform distribution on $\mathcal{X}$, and whose transition probabilities are integer multiples of $1/Q$, for some integer $Q$. I show how a simulation of $n$ transitions of this chain starting at $x_0$ can be viewed as applying a random permutation on the space $\mathcal{X} \times \mathcal{U}$, where $\mathcal{U} = \{0, 1, \ldots, Q-1\}$, to the start state $(x_0, u_0)$, with $u_0$ drawn uniformly from $\mathcal{U}$. This result can be applied to a non-uniform distribution with probabilities that are integer multiples of $1/P$, for some integer $P$, by representing it as the marginal distribution for $\mathcal{X}$ from the uniform distribution on a suitably-defined subset of $\mathcal{X} \times \mathcal{Y}$, where $\mathcal{Y} = \{0, 1, \ldots, P-1\}$. By letting $Q$, $P$, and the cardinality of $\mathcal{X}$ go to infinity, this result can be generalized to non-rational probabilities and to continuous state spaces, with permutations on a finite space replaced by volume-preserving one-to-one maps from a continuous space to itself. These constructions can be efficiently implemented for chains commonly used in Markov chain Monte Carlo (MCMC) simulations. I present two applications in this context — simulation of $K$ realizations of a chain from $K$ initial states, but with transitions defined by a single stream of random numbers, as may be efficient with a vector processor or multiple processors, and use of MCMC to improve an importance sampling distribution that already has substantial overlap with the distribution of interest. I also discuss the implications of this "permutation MCMC" method regarding the role of randomness in MCMC simulation, and the potential use of non-random and quasi-random numbers.

## 1 Introduction

Markov chain Monte Carlo (MCMC) simulation might seem to be a fundamentally contractive process. A simulation started from a broad initial distribution must, after many transitions, be concentrated in the possibly much-smaller region that has high probability under the equilibrium distribution being sampled. This implies that the random map determined by the random numbers underlying the Markov chain transitions must be contractive — for a finite state space, it must map a large set of states to a smaller set of states, and for a continuous state space, it must map a set of large volume to a set of smaller volume. This contractive property underlies the ability to "couple" a set of chains so that eventually they all "coalesce" to the same state, as is exploited by methods such as coupling from the past (Propp and Wilson, 1996) and circular coupling (Neal, 1999/2002).

In this paper, I show that with a simple extension of the state space this contractive behaviour can be converted to a non-contractive map — which is a permutation when this extended state space is finite, or a one-to-one map that preserves volume when the extended state space is continuous. This result was suggested by the volume-preserving property of Hamiltonian dynamics (see Neal, 2010), which can be used to define an importance sampling procedure based on annealing (Neal, 2005). I expect that the result in this paper can be applied to produce similar procedures that combine annealing and importance sampling using other MCMC techniques. However, I will leave that for future work, and instead present two simpler applications of what I will call "permutation MCMC".

One application is to parallel simulation from many initial states using a single stream of random numbers. Using a single random number stream for all parallel chains may reduce the computational cost, perhaps especially if the parallelism takes the form of vector operations. (At worst, it costs the same as using multiple streams, if due to high communication cost it is fastest to compute the same stream separately in each processor.) Using a single stream also avoids the issue of how to set up multiple streams that are unrelated, a problem that is discussed, for example, by Wu and Huang (2006).

However, some ways of using a single random number stream to define transitions in parallel chains lead to the chains coalescing to the same state, or to states that approach each other increasingly closely, eliminating the benefit of multiple chains in producing better estimates. I will demonstrate that this is avoided when transitions are defined as random permutations, or as random volume-preserving maps.

A second application is to improving importance sampling, in which expectations with respect to some distribution of interest are found using points drawn from some approximating distribution that is easier to sample from. This method produces good results only when the importance sampling distribution is a sufficiently good approximation, and most crucially does not give very low probability to regions that have significant probability under the distribution of interest. This can be hard to guarantee in high-dimensional problems. We might improve an importance sampling distribution that is close to being adequate — in the sense that it at least has substantial overlap with the distribution of interest — by performing some number of transitions of a Markov chain that leaves the distribution of interest invariant starting from a point drawn from the original importance sampling distribution. This will improve the approximation even if the random numbers used to simulate this Markov chain are fixed, provided we choose the number of transitions randomly, so that the final importance sampling distribution is a mixture of distributions after varying numbers of transitions. With standard methods of simulation, however, computing the importance sampling probabilities (or densities), as needed to find appropriate weights, will often be infeasible, because the same final point might be produced from several initial points (or for a continuous distribution, an unknown change in volume may alter the densities). I will show how this problem can be bypassed by viewing the transitions as applying a random permutation (or volume-preserving map), for which the probability (or density) of the final point is the same as that of the initial point.

Recently, Murray and Elliott (2012) independently devised MCMC simulation methods equivalent or similar to some of the methods I present below, though without additional variables needed to produce a volume-preserving map. Their aim was to find a simulation method that is insensitive to dependence in the stream of random numbers used, or even to whether they are actually random. I conclude this paper by also discussing what permutation MCMC says about the role of randomness, and how MCMC efficiency might be improved by using permutation MCMC with non-random or quasi-random numbers.

The programs and scripts used for the experiments in this paper are available from my web page.

# 2 Viewing MCMC for a uniform distribution as a random permutation

I will begin with the simple case of a Markov chain that samples from the uniform distribution on some finite state space. In the following sections, I generalize to other discrete and continuous distributions.

Consider a Markov chain on some finite state space, $\mathcal{X}$, which we can take to be $\{0, \ldots, M-1\}$. Let the probability of this chain transitioning to state $x'$ when the current state is $x$ be $T(x, x')$, and for the moment assume these transition probabilities are integer multiples of $1/Q$, for some integer $Q$. We wish to use this chain to sample from the uniform distribution on $\mathcal{X}$, so $T$ will be chosen to leave this uniform distribution invariant — that is,

$$\sum_{x \in \mathcal{X}} (1/M) \, T(x, x') \;=\; 1/M \tag{1}$$

If we view $T$ as a matrix, this condition is equivalent to all its columns (as well as all its rows) summing to one.

A standard way to simulate a realization, $x_0, x_1, x_2, \ldots$ of this chain, starting from some state $x_0$, is to draw $u_0, u_1, u_2, \ldots$ independently from the uniform distribution on $\mathcal{U} = \{0, 1, \ldots, Q-1\}$ and then set

$$x_{i+1} \;=\; \max\left\{ x' : Q \sum_{x=0}^{x'-1} T(x_i, x) \leq u_i \right\} \tag{2}$$

The first step in converting this simulation to a random permutation is to extend the state space to $\mathcal{X} \times \mathcal{U}$. We then draw a value for $u_0$ uniformly from $\mathcal{U}$. Subsequent transitions from $(x_0, u_0)$ are defined using $s_0, s_1, s_2, \ldots$, which are independently drawn uniformly from $\mathcal{U}$. (We will see below that $s_0, \ldots, s_n$ specify a random permutation mapping $(x_0, u_0)$ to $(x_n, u_n)$.) From the state $(x_i, u_i)$, $x_{i+1}$ is derived from $x_i$ and $u_i$ as in equation (2) above, and $u_{i+1}$ is derived from $x_i$, $u_i$, $s_i$, and $x_{i+1}$ as follows:

$$u_{i+1} \;=\; s_i \,+\, u_i \,-\, Q\sum_{x=0}^{x_{i+1}-1} T(x_i, x) \,+\, Q \sum_{x=0}^{x_i - 1} \widetilde{T}(x_{i+1}, x) \pmod{Q} \tag{3}$$

where $\widetilde{T}(x, x') = T(x', x)$ are the transition probabilities for the reversed chain.

To see informally the rationale for this, note that the terms on the right other than $s_i$ define a value for $u$ that would lead back to $x_i$ if an equation analogous to (2) were applied with $T$ replaced by $\widetilde{T}$. This part of the map from $(x_i, u_i)$ to $(x_{i+1}, u_{i+1})$ is therefore a permutation. Adding $s_i$ modulo $Q$ is a random circular permutation, so the full map from $(x_i, u_i)$ to $(x_{i+1}, u_{i+1})$ is a random permutation as well. For any $n > 0$, the map from $(x_0, u_0)$ to $(x_n, u_n)$, being a composition of random permutations, is also a random permutation.

Furthermore, if we look at only a single realization of the chain, setting $u_{i+1}$ to an independent random $s_i$ plus anything (mod $Q$) has the same effect as setting $u_{i+1}$ independently at random, so the joint distribution of $(x_1, u_1)$, $(x_2, u_2)$, ... is the same as for the standard method of simulation.

Appendix A shows in detail that the map $(x_i, u_i) \to (x_{i+1}, u_{i+1})$ defined by equations (2) and (3) is a permutation, by explicitly exhibiting the inverse map.

Here is a matrix of transition probabilities for a simple example with $M = 4$ states:

$$T = \begin{bmatrix} 2/3 & 1/3 & 0 & 0 \\ 1/3 & 1/3 & 1/3 & 0 \\ 0 & 1/3 & 1/3 & 1/3 \\ 0 & 0 & 1/3 & 2/3 \end{bmatrix} \quad (4)$$

All the columns above sum to one, so these transitions leave the uniform distribution on $\mathcal{X} = \{0, 1, 2, 3, 4\}$ invariant. Since all the transition probabilities are multiples of $1/3$, we can set $Q = 3$, and hence $\mathcal{U} = \{0, 1, 2\}$. Note that these transitions are reversible — that is, $\widetilde{T}(x, x') = T(x', x) = T(x, x')$.

The permutation maps from $(x_i, u_i)$ to $(x_{i+1}, u_{i+1})$ when $s_i$ has each of its possible values are shown here:

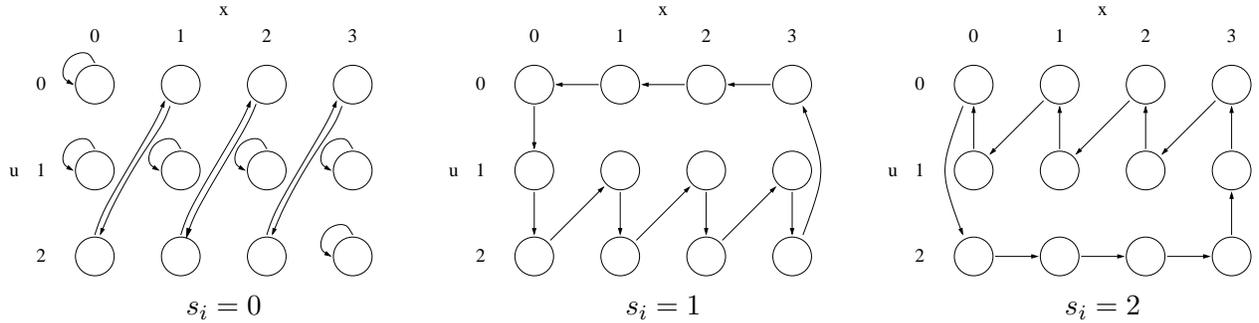

In these diagrams, the array of circles represents all possible $(x, u)$ pairs, and the arrows show how such a pair for $(x_i, u_i)$ is mapped to $(x_{i+1}, u_{i+1})$. For example, the arrow out of the state with $x_i = 1$ and $u_i = 2$ goes to a state with $x_{i+1} = 2$, regardless of the value of $s_i$, since $Q\,(T(1,0)+T(1,1)) = 3(2/3) = 2 \leq 2$, so the maximum in equation (2) will be $x' = 2$. When $s_i = 0$, equation (3) will find a value for $u_{i+1}$ that would lead back to the state $x_i = 1$ starting from $x_{i+1} = 2$, which requires that $u_{i+1}$ be at least $Q\,T(2,0) = 0$, to which must be added the amount by which $u_i$ was greater than the minimum needed for the transition to $x_{i+1} = 2$ to be taken (essential to avoid two states mapping to the same new state), which in this case is 0. The result is that the diagram for $s_i = 0$ has the transition $(1, 2) \to (2, 0)$.

The maps for $s_i \neq 0$ can be obtained from the map for $s_i = 0$ by circularly shifting the $u_{i+1}$. Note that when, as here, the transitions are reversible, the diagram for $s_i = 0$ will consist entirely of single states with arrows pointing to themselves and pairs of states connected by arrows both ways.

For comparison, here are the maps produced by the $T$ defined in equation (4) when equation (3) is replaced by $u_{i+1} = s_i + u_i \pmod{Q}$:

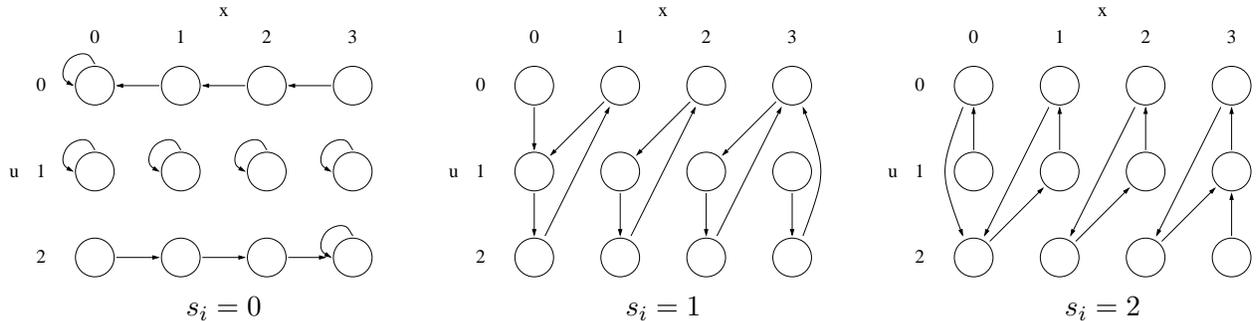

Some states have zero or two incoming arrows, so these are clearly not permutations.

Below, are the transition probabilities, $T$, for a non-reversible Markov chain with $M = 4$ states that leaves the uniform distribution on $\mathcal{X} = \{0, 1, 2, 3, 4\}$ invariant, along with the reverse transition probabilities, $\widetilde{T}$, found by transposing $T$:

$$T \;=\; \begin{bmatrix} 1/2 & 1/2 & 0 & 0 \\ 1/4 & 1/4 & 1/4 & 1/4 \\ 0 & 0 & 1/2 & 1/2 \\ 1/4 & 1/4 & 1/4 & 1/4 \end{bmatrix}, \qquad \widetilde{T} \;=\; \begin{bmatrix} 1/2 & 1/4 & 0 & 1/4 \\ 1/2 & 1/4 & 0 & 1/4 \\ 0 & 1/4 & 1/2 & 1/4 \\ 0 & 1/4 & 1/2 & 1/4 \end{bmatrix} \qquad (5)$$

Since all transition probabilities are multiples of $1/4$, we can set $Q = 4$, so that $\mathcal{U} = \{0, 1, 2, 3\}$. The permutation maps for this example from $(x_i, u_i)$ to $(x_{i+1}, u_{i+1})$ for each $s_i$ are as follows:

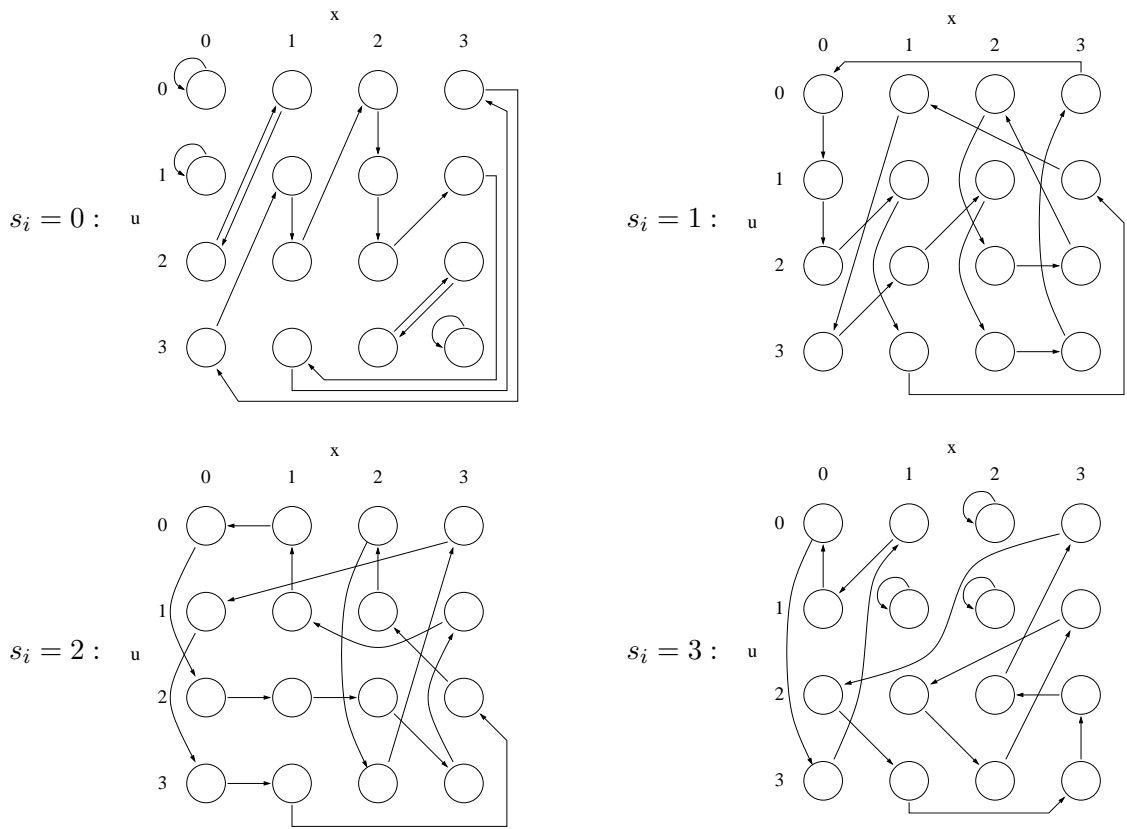

In this example, the value of $x_{i+1}$ that follows $(x_i, u_i)$ is determined using equation (2) in the same way as for a reversible chain, but the value of $u_{i+1}$ when $s_i = 0$ is not one that would lead back to $x_i$ if $T$ were applied starting from $x_{i+1}$, but is rather a value that would lead back to $x_i$ if the reverse transition, $\widetilde{T}$, were applied.

In real MCMC applications, unlike these examples, the state space is enormous, and transition probabilities are defined algorithmically, rather than via an explicit table. One may then ask whether the computation of $x_{i+1}$ and $u_{i+1}$ from $x_i$ and $u_i$ according to equations (2) and (3) is feasible. I will defer consideration of this issue to the following sections, in which the method is generalized to non-uniform distributions and to continuous state spaces.

## 3 Generalization to non-uniform discrete distributions

As a first step in generalizing the result in the previous section, let us consider a distribution on $\mathcal{X} = \{0, \ldots, M-1\}$ with probabilities proportional to a function $\pi(x)$ whose values are all integer multiples of $1/P$, for some positive integer $P$. (We may not know the constant of proportionality, $1/\sum_x \pi(x)$.) This distribution can be obtained as the marginal distribution on $\mathcal{X}$ obtained from a uniform joint distribution on the following subset of $\mathcal{X} \times \mathcal{Y}$, where $\mathcal{Y} = \{0, \ldots, P-1\}$:

$$\mathcal{Z} \;=\; \{\,(x,y) \,:\, 0 \leq y < P\pi(x)\,\} \tag{6}$$

The cardinality of $\mathcal{Z}$ is $M^+ = P\sum_x \pi(x)$.

This construction is analogous to what is done for "slice sampling" MCMC methods (Neal, 2003), in which Markov transitions are defined on this extended state space. Here, I will assume that our MCMC method is defined in terms of transitions on $\mathcal{X}$, with the introduction of the extended space $\mathcal{X} \times \mathcal{Y}$ being only a device to allow these transitions to be expressed as permutations. (However, transitions defined on $\mathcal{X} \times \mathcal{Y}$ could be accommodated if desired.)

Suppose that we have defined a Markov chain on $\mathcal{X}$, with transition probabilities $T(x, x')$, all integer multiples of $1/Q$, that leaves the distribution $\pi(x)$ invariant. We can define a Markov chain on $\mathcal{Z}$ that leaves the uniform distribution on $\mathcal{Z}$ invariant, with transition probabilities as follows:

$$T^+((x,y),(x',y')) \;=\; \begin{cases} \dfrac{T(x,x')}{P\pi(x')} & \text{if } 0 \leq y' < P\pi(x') \\ 0 & \text{otherwise} \end{cases} \tag{7}$$

These transition probabilities are all integer multiples of $1/Q^+$, where $Q^+ = (\max_x P\pi(x))!\, Q$.

To confirm that $T^+$ leaves the uniform distribution invariant, note that for any $(x', y')$ in $\mathcal{Z}$,

$$\sum_{(x,y)\in\mathcal{Z}} \frac{1}{M^+}\, T^+((x,y),(x',y')) \;=\; \frac{1}{M^+} \sum_x P\pi(x)\, \frac{T(x,x')}{P\pi(x')} \;=\; \frac{1}{M^+}\frac{1}{\pi(x')}\sum_x \pi(x) T(x,x') \;=\; \frac{1}{M^+} \tag{8}$$

The effect of applying the original transitions, $T$, starting from some initial state, $x_0$, can be duplicated by drawing $y_0$ uniformly from $\{0, \ldots, P\pi(x_0)\}$ and then using the transitions $T^+$ to simulate states $(x_1, y_1), (x_2, y_2), \ldots$ The resulting distribution for $x_1, x_2, \ldots$ is the same as if $T$ were applied starting with $x_0$ — the transition from $(x_i, y_i)$ to $(x_{i+1}, y_{i+1})$ defined by $T^+$ ignores $y_i$, and gives equal probabilities of $T(x_i, x_{i+1})/P\pi(x_{i+1})$ to $P\pi(x_{i+1})$ values of $y'$, so the total probability for a value $x_{i+1}$ to follow $x_i$ is $T(x_i, x_{i+1})$.

An MCMC simulation using $T$ that samples from $\mathcal{X}$ with probabilities given by $\pi$ can therefore be replaced by a simulation using $T^+$ that samples from $\mathcal{Z}$ with uniform probabilities. This simulation on the extended state space $\mathcal{Z}$ can be expressed as a random permutation, as described in Section 2. To do this, we must decide on an ordering of states in $Z$. In this paper, I will use a lexicographical order (first on $x$, then on $y$), in which an $(x, y)$ pair in $\mathcal{Z}$ is associated with a label, $x^+$, in $\mathcal{X}^+ = \{0, 1, \ldots, M^+ - 1\}$, according to the following map:

$$X^+(x,y) \;=\; y \;+\; P\sum_{\ddot{x}=0}^{x-1} \pi(\ddot{x}) \tag{9}$$

The inverse of this map takes $x^+$ to $(X(x^+), Y(x^+))$, where

$$X(x^+) = \max\left\{x : P\sum_{\ddot{x}=0}^{x-1} \pi(\ddot{x}) \leq x^+\right\}, \quad Y(x^+) = x^+ - P\sum_{\ddot{x}=0}^{X(x^+)-1} \pi(\ddot{x}) \quad (10)$$

We can now apply the method of Section 2, replacing references there to $\mathcal{X}$, $M$, $T$, and $Q$ with references to $\mathcal{X}^+$, $M^+$, $T^+$, and $Q^+$, after redefining $T^+$ to apply to states in $\mathcal{X}^+$ rather than the associated pairs in $\mathcal{Z}$. In terms of $T^+$ as redefined in this way, the transitions defined by equation (2) and equation (3) become

$$x_{i+1}^+ = \max\left\{x^{+\prime} : Q^+ \sum_{x^+=0}^{x^{+\prime}-1} T^+(x_i^+, x^+) \leq u_i\right\} \quad (11)$$

$$u_{i+1} = s_i + u_i - Q^+ \sum_{x^+=0}^{x_{i+1}^+-1} T^+(x_i^+, x^+) + Q^+ \sum_{x^+=0}^{x_i^+-1} \widetilde{T}^+(x_{i+1}^+, x^+) \pmod{Q^+} \quad (12)$$

where $\widetilde{T}^+(x, x') = T^+(x', x)$.

It is convenient to re-express these transitions in terms of $(x, y)$ and the original transitions $T$. Noting that $X^+(x,y) \geq X^+(x',y)$ when $x \geq x'$ and $X(x^+) \geq X(x^{+\prime})$ when $x^+ \geq x^{+\prime}$, we can write one of the sums appearing above as follows, with $x_i = X(x_i^+)$, $x_{i+1} = X(x_{i+1}^+)$, and $y_{i+1} = Y(x_{i+1}^+)$:

$$Q^+ \sum_{x^+=0}^{x_{i+1}^+-1} T^+(x_i^+, x^+) = Q^+\left[\sum_{x=0}^{x_{i+1}-1} \sum_{y=0}^{P\pi(x)} T^+(x_i^+, X^+(x,y)) + \sum_{y=0}^{y_{i+1}-1} T^+(x_i^+, X^+(x_{i+1}, y))\right] \quad (13)$$

$$= Q^+\left[\sum_{x=0}^{x_{i+1}-1} T(x_i, x) + \frac{y_{i+1}}{P\pi(x_{i+1})} T(x_i, x_{i+1})\right] \quad (14)$$

To rewrite the other sum above, we let $y_i = Y(x_i^+)$, and define the reverse transition probabilities for the original chain as $\widetilde{T}(x, x') = T(x', x)\pi(x')/\pi(x)$. We then have

$$Q^+ \sum_{x^+=0}^{x_i^+-1} \widetilde{T}^+(x_{i+1}^+, x^+) = Q^+\left[\sum_{x=0}^{x_i-1} \widetilde{T}(x_{i+1}, x) + \frac{y_i}{P\pi(x_i)} \widetilde{T}(x_{i+1}, x_i)\right] \quad (15)$$

Also, note that $y_i/P\pi(x_i)$ and $y_{i+1}/P\pi(x_{i+1})$ are less than one.

The transitions of equations (11) and (12) can now be rewritten using expressions (14) and (15) for the sums they contain, as follows:

$$x_{i+1} = \max\left\{x' : Q^+ \sum_{x=0}^{x'-1} T(x_i, x) \leq u_i\right\} \quad (16)$$

$$y_{i+1} = \left\lfloor P\pi(x_{i+1})\left(u_i - Q^+ \sum_{x=0}^{x_{i+1}-1} T(x_i, x)\right) \Big/ (Q^+ T(x_i, x_{i+1}))\right\rfloor \quad (17)$$

$$u_{i+1} = s_i + u_i - Q^+ \sum_{x=0}^{x_{i+1}-1} T(x_i, x) - Q^+ T(x_i, x_{i+1}) \frac{y_{i+1}}{P\pi(x_{i+1})}$$

$$+ Q^+ \sum_{x=0}^{x_i-1} \widetilde{T}(x_{i+1}, x) + Q^+ \widetilde{T}(x_{i+1}, x_i) \frac{y_i}{P\pi(x_i)} \pmod{Q^+} \quad (18)$$

Defining $y^* = y/P$, so that $y^*$ is in $[0, \pi(x))$, and letting $u^* = u/Q^+$ and $s^* = s/Q^+$, both in $[0, 1)$, we can rewrite the above equations as follows:

$$x_{i+1} = \max\left\{x' : \sum_{x=0}^{x'-1} T(x_i, x) \leq u_i^*\right\} \tag{19}$$

$$y_{i+1}^* = \left\lfloor P\pi(x_{i+1})\left(u_i^* - \sum_{x=0}^{x_{i+1}-1} T(x_i, x)\right) \Big/ T(x_i, x_{i+1})\right\rfloor \Big/ P \tag{20}$$

$$u_{i+1}^* = s_i^* + u_i^* - \sum_{x=0}^{x_{i+1}-1} T(x_i, x) - T(x_i, x_{i+1}) \frac{y_{i+1}^*}{\pi(x_{i+1})}$$

$$+ \sum_{x=0}^{x_i-1} \widetilde{T}(x_{i+1}, x) + \widetilde{T}(x_{i+1}, x_i) \frac{y_i^*}{\pi(x_i)} \pmod 1 \tag{21}$$

where $U(\text{mod } 1)$ means $U - \lfloor U \rfloor$, the fractional part of $U$.

If we now let $P$ (and hence also $M^+$ and $Q^+$) go to infinity, which corresponds to letting the probabilities $\pi(x)$ take on any real values in $[0, 1]$, we get a simpler expression for $y_{i+1}^*$, which when substituted into equation (21) gives a simpler expression for $u_{i+1}^*$. The final result is the following transition:

$$x_{i+1} = \max\left\{x' : \sum_{x=0}^{x'-1} T(x_i, x) \leq u_i^*\right\} \tag{22}$$

$$y_{i+1}^* = \pi(x_{i+1})\left(u_i^* - \sum_{x=0}^{x_{i+1}-1} T(x_i, x)\right) \Big/ T(x_i, x_{i+1}) \tag{23}$$

$$u_{i+1}^* = s_i^* + \sum_{x=0}^{x_i-1} \widetilde{T}(x_{i+1}, x) + \widetilde{T}(x_{i+1}, x_i) \frac{y_i^*}{\pi(x_i)} \pmod 1 \tag{24}$$

In this limit, the values $s_0^*, s_1^*, s_2^*, \ldots$ that determine the random transitions are drawn independently from the uniform distribution on $[0, 1)$. Note that this transition preserves the property that $y_{i+1}^*$ is in $[0, \pi(x_{i+1}))$. In a program, it may be more efficient to maintain the quantity $y^*/\pi(x)$ rather than $y^*$.

Since the map defined by equations (22) to (24) is a limit of discrete permutation maps, for an increasingly fine uniform grid, it should not only be one-to-one, but also preserve volume. This is shown in Appendix B, by explicitly exhibiting the inverse of the map, and showing that its continuous part has a Jacobian matrix whose determinant has absolute value one. Note, however, that a computer implementation of this map that uses floating-point representations of $y^*$ and $u^*$ may not be exactly reversible, or may not exactly preserve volume, due to round-off error.

This map is similar to one defined by Murray and Elliott (2012), in their equations (9), (10), and (17). However, they have no equivalent of $y^*$, and their update for $u^*$ is what would be obtained by replacing $y_i^*$ in equation (24) above by the definition of $y_{i+1}^*$ from equation (23). Because of this difference, Murray and Elliott's map does not preserve volume. This may not matter for their purpose of producing an MCMC method that works with dependent random number streams.

As an example of the map defined by equations (22) to (24), consider a state space of $\mathcal{X} = \{0, 1, 2\}$ with probabilities $\pi(0) = 3/10$, $\pi(1) = 1/10$, and $\pi(2) = 6/10$. Let the transition probabilities, $T$, and their reversal, $\widetilde{T}$, be given by the following matrices:

$$T = \begin{bmatrix} 1/3 & 1/3 & 1/3 \\ 0 & 0 & 1 \\ 1/3 & 0 & 2/3 \end{bmatrix}, \quad \widetilde{T} = \begin{bmatrix} 1/3 & 0 & 2/3 \\ 1 & 0 & 0 \\ 1/6 & 1/6 & 2/3 \end{bmatrix} \qquad (25)$$

The diagram below shows how $(x_i, y_i^*, u_i^*)$ is mapped to $(x_{i+1}, y_{i+1}^*, u_{i+1}^*)$ when $s_i = 0$:

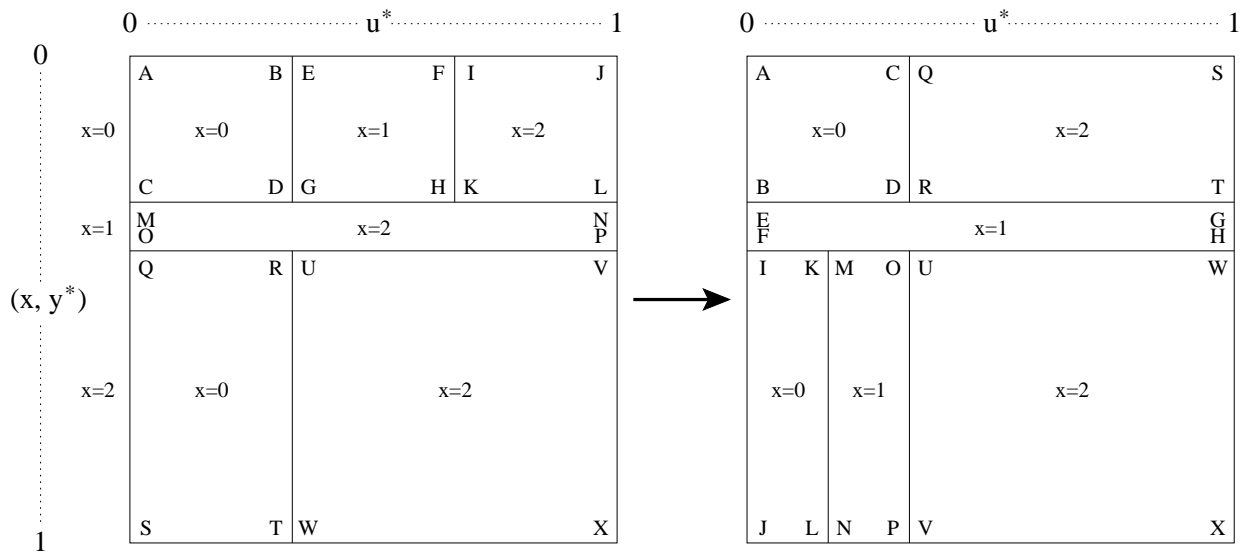

The horizontally dimension of each of the two squares above represents the range of $[0, 1)$ for $u^*$. The vertical dimension represents the range of $(x, y^*)$, with $[0, 1)$ divided into a section for each value of $x$ of size $\pi(x)$, with the value of $y^*$ going from 0 to $\pi(x)$ within each such section. Rectangles in the square on the left correspond to transition probabilities, $T(x_i, x_{i+1})$. Points in such a rectangle are mapped to points in a rectangle of equal area in the square on the right, as identified by the letters labelling the corners. Such rectangles on the right correspond to reverse transition probabilities, $\widetilde{T}(x_{i+1}, x_i)$.

If $\pi(x)$ is only proportional to the probabilities, not equal, the the vertical scale in the diagram above would be stretched or compressed by the factor $\sum_x \pi(x)$.

Often, the update defined by equations (22) to (24) will be implementable nearly as efficiently as a standard update for the same transition probabilities, in which just equation (22) is used with a value for $u_i^*$ drawn uniformly from $[0, 1)$. If the search over $x'$ involving the sum in equation (22) can be done efficiently, it will usually deliver the value of this sum for $x_{i+1}$ at little or no extra cost, as needed for equation (23). The sum in equation (24) involving the reversed transition probabilities will typically be feasible if the analogous sum for the original transition probabilities is feasible, certainly so if the chain is reversible, so that the original and reverse transition probabilities are the same.

In one common class of problems, $\mathcal{X} = \mathcal{X}_1 \times \cdots \times \mathcal{X}_d$, so that a state $x$ is composed of values for $d$ discrete variables, with the number of possible values for each of these variables (ie, the cardinality of each $\mathcal{X}_k$) being small. One popular approach to sampling a distribution on such a state space is to simulate a Markov chain that applies $d$ updates in succession, each of which changes only one component of $x$. Examples of such problems are the Ising and Potts models of statistical physics, and Bayesian mixture models with conjugate priors in which the continuous parameters have been integrated out, leaving only discrete class indicators for each data point (as in Algorithm 3 in (Neal, 2000)).

Denoting the transition probabilities when only component $k$ of the state is updated by $T_k$, we see that for problems of this sort $T_k(x, x')$ will be non-zero for only a small number of values for $x'$ (at most the cardinality of $\mathcal{X}_k$), even when the cardinality of $\mathcal{X}$ is enormous. Furthermore, the values of $x'$ for which $T_k(x, x')$ may be non-zero are easily identified, so the sums in equations (22) to (24) can be efficiently computed by explicit summation.

The Metropolis-Hastings algorithm (Metropolis, *et al*, 1953; Hastings, 1970) is another popular way to define a Markov chain on $\mathcal{X}$ that converges to the distribution defined by $\pi(x)$. This method draws a proposal, $\hat{x}_i$, for the state to follow $x_i$ according to some proposal distribution with probabilities given by $S(x_i, \hat{x}_i)$, and then accepts this proposal as $x_{i+1}$ with probability $a(x_i, \hat{x}_i)$, where

$$a(x, \hat{x}) \;=\; \min\left[1, \frac{\pi(\hat{x})\, S(\hat{x}, x)}{\pi(x)\, S(x, \hat{x})}\right] \tag{26}$$

If the proposal is not accepted, $x_{i+1} = x_i$. The resulting transition probabilities are reversible, and are given by

$$T(x, x') \;=\; \begin{cases} S(x, x')\, a(x, x') & \text{if } x \neq x' \\ S(x, x) \;+\; \sum_{\ddot{x} \in \mathcal{X}} S(x, \ddot{x})\,(1-a(x, \ddot{x})) & \text{if } x = x' \end{cases} \tag{27}$$

Standard simulation of such a chain is feasible if one can sample from $S(x, \hat{x})$ and compute $\pi(x)$ and $S(x, \hat{x})$ (with the latter not required if $S(x, \hat{x}) = S(\hat{x}, x)$ so that their ratio in the acceptance probability is always one). Simulation using equations (22) to (24) might be more difficult, however, because computation of $T(x, x)$ involves the rejection probabilities (whose computation requires evaluating $\pi$) for all possible proposals.

This problem can be bypassed by changing how $u_i^*$ is used to determine $x_{i+1}$. In equation (22), the full range of $u_i^*$ is partitioned into contiguous sub-intervals associated with each value for $x_{i+1}$. Instead, we can partition the range of $u_i^*$ into subintervals corresponding to proposals, with sizes given by $S(x_i, \hat{x}_i)$, and then further subdivide each of these subintervals into a part associated with acceptance of the proposal and a part (possibly empty) associated with rejection. A self-transition, with probability $T(x, x)$, is then represented by the union of an interval of size $S(x, x)$, corresponding to proposing (and accepting) the same state as the current state, and zero or more intervals associated with rejection of proposals to move to a different state.

In detail, a Metropolis-Hastings transition can be performed as follows:

$$\hat{x}_{i+1} \;=\; \max\left\{x' : \sum_{x=0}^{x'-1} S(x_i, x) \;\leq\; u_i^*\right\} \tag{28}$$

$$a_{i+1} \;=\; \left(u_i^* - \sum_{x=0}^{\hat{x}_{i+1}-1} S(x_i, x)\right) \bigg/ S(x_i, \hat{x}_{i+1}) \tag{29}$$

$$x_{i+1} \;=\; \begin{cases} \hat{x}_{i+1} & \text{if } a_{i+1} < a(x_i, \hat{x}_{i+1}) \\ x_i & \text{otherwise} \end{cases} \tag{30}$$

$$y_{i+1}^* \;=\; \begin{cases} \pi(x_{i+1})\, a_{i+1}\, /\, a(x_i, \hat{x}_{i+1}) & \text{if } a_{i+1} < a(x_i, \hat{x}_{i+1}) \\ y_i^* & \text{otherwise} \end{cases} \tag{31}$$

$$u^*_{i+1} = \begin{cases} s^*_i + \sum_{x=0}^{x_i-1} S(x_{i+1}, x) + S(x_{i+1}, x_i) \, a(x_{i+1}, x_i) \, \dfrac{y^*_i}{\pi(x_i)} \pmod{1} & \text{if } a_{i+1} < a(x_i, \hat{x}_{i+1}) \\ s^*_i + u^*_i \pmod{1} & \text{otherwise} \end{cases} \quad (32)$$

Appendix C shows that this map is indeed one-to-one and volume preserving.

As an example, consider Metropolis-Hastings transitions that leave invariant the distribution on $\mathcal{X} = \{0, 1, 2, 3\}$ that has probabilities

$$\pi(0) = 1/3, \quad \pi(1) = 1/3, \quad \pi(2) = 2/9, \quad \pi(3) = 1/9 \qquad (33)$$

using the matrix of proposal probabilities, $S(x, \hat{x})$, on the left below, which produces the transition probabilities, $T(x, x')$, shown on the right:

$$S = \begin{bmatrix} 1/2 & 1/2 & 0 & 0 \\ 1/3 & 1/3 & 1/3 & 0 \\ 0 & 1/3 & 1/3 & 1/3 \\ 0 & 0 & 1/2 & 1/2 \end{bmatrix}, \quad T = \begin{bmatrix} 1/2 + 1/6 & 1/3 & 0 & 0 \\ 1/3 & 1/3 + 1/9 & 2/9 & 0 \\ 0 & 1/3 & 1/3 + 1/12 & 1/4 \\ 0 & 0 & 1/2 & 1/2 \end{bmatrix} \qquad (34)$$

The resulting map from $(x_i, y^*_i, u^*_i)$ to $(x_{i+1}, y^*_{i+1}, u^*_{i+1})$ defined by equations (29) to (32), when $s_i = 0$, is pictured below:

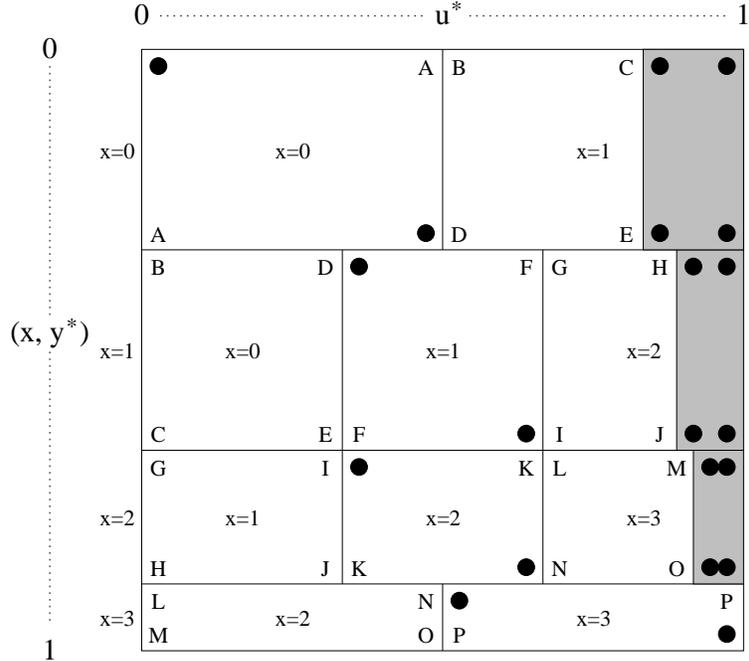

Since Metropolis-Hastings transitions are reversible, this one diagram shows both the transition and its reversal. A corner of a rectangle marked by a black dot is not moved by the map; a corner labelled by a letter is mapped to the other corner with the same label. The shaded rectangles are associated with rejected proposals.

Implementing a Metropolis-Hastings transition in this fashion will often be feasible, since the sums in equations (28), (29), and (32) involve only the proposal probabilities, which will often have a tractable form. However, if such an implementation is still costly, an alternative is to express the proposal distribution as a mixture of simpler proposal distributions, as follows:

$$S(x, \hat{x}) = \sum_{\delta} p(\delta) S_{\delta}(x, \hat{x}) \tag{35}$$

Here, $p(\delta)$ are the probabilities for some set of simpler proposal distributions, $S_\delta$. Rather than use $S$ itself as a proposal distribution, we perform transition $i$ by randomly selecting a value $\delta_i$ according to $p(\delta)$ and then performing a Metropolis-Hastings transition with $S_{\delta_i}$ as the proposal distribution. (Note that it is generally valid to choose from a set of transition probabilities randomly at each transition of the chain, as long as each transition leaves $\pi$ invariant, and the choice is made independently of other choices.) We consider the sequence $\delta_0, \delta_1, \delta_2, \ldots$ to be part of the specification of the random volume-preserving map (along with $s_0^*, s_1^*, s_2^*, \ldots$).

For example, a "random walk" Metropolis transition can be defined when $\mathcal{X}$ has a group structure, with operations $\oplus$ and $\ominus$. We can then let $\delta$ be an element of this group, and define

$$S_\delta(x, \hat{x}) = \begin{cases} 1/2 & \text{if } \hat{x} = x \oplus \delta \\ 1/2 & \text{if } \hat{x} = x \ominus \delta \\ 0 & \text{otherwise} \end{cases} \tag{36}$$

This is symmetric, with $S_\delta(x, \hat{x}) = S_\delta(\hat{x}, x)$, so the acceptance probability of equation (26) simplifies to $\min[1, \pi(\hat{x})/\pi(x)]$. It is easy to implement equations (28) to (32) when $S$ is set to such an $S_\delta$. The probabilities $p(\delta)$ will control the sizes and directions of jumps that are proposed.

## 4 Generalization to continuous distributions

I will begin generalizing to continuous distributions by considering a Markov chain on a state space that is an interval of real numbers, which will be denoted as $\mathcal{X}^* = (a, b)$. I will assume that the chain leaves invariant a distribution that has a density function proportional to $\pi^*(x^*)$, and that the transitions of the chain can be expressed as conditional density functions, with $T^*(x^*, x^{*\prime})$ the conditional density for the state $x^{*\prime}$ to follow the state $x^*$. I also assume that the total variation distance between the transition density functions $T^*(x^*, \cdot)$ and $T^*(x^* + \epsilon, \cdot)$ goes to zero as $\epsilon$ goes to zero — that is, a slight perturbation in the current state has only a small probability of altering what state follows — except perhaps at a finite set of values for $x^*$.

A chain with a finite state space $\mathcal{X} = \{0, \ldots, M-1\}$ can approximate the above chain arbitrarily well as $M$ goes to infinity. We associate an $x^* \in \mathcal{X}^*$ with each $x \in \mathcal{X}$ according to $x^* = a + (x+1/2)h$, where $h = (b-a)/M$, and define the transition probabilities for the approximating chain by

$$T(x, x') = \int_{a+hx'}^{a+h(x'+1)} T^*(a + (x + 1/2)h, \, x^*) \, dx^* \tag{37}$$

This corresponds more-or-less to how simulations of Markov chains for continuous state spaces are actually done on computers, which can represent numbers only to some finite precision.

We can define a distribution on $\mathcal{X}$ with probabilities proportional to $\pi(x) = \pi^*(a + (x+1/2)h)$. For any finite $M$, the transitions of equation (37) will in general not leave this distribution exactly invariant.

However, we can nevertheless use $\pi(x)$ to define transitions on an extended space using equations (22) to (24). In the limit as $M$ increases and hence $h$ approaches zero, we can write the transition probabilities as $T(x, x') \approx h\, T^*(a + (x + 1/2)h,\, a + (x' + 1/2)h) = T^*(x^*, x^{*\prime})$. Equations (22) and (24) can then be written as follows:

$$x^*_{i+1} = \max\left\{ x^{*\prime} : \int_a^{x^{*\prime}} T^*(x^*_i, x^*)\, \leq u^*_i \right\} = F_*^{-1}(x^*_i, u^*_i) \tag{38}$$

$$u^*_{i+1} = s^*_i + \int_a^{x^*_i} \widetilde{T}^*(x^*_{i+1}, x^*)\, dx^* \pmod{1} = s^*_i + \widetilde{F}_*(x^*_{i+1}, x^*_i) \pmod{1} \tag{39}$$

Here, $F_*(x, x') = \int_a^{x'} T^*(x, x')\, dx'$ gives the cumulative distribution functions for the transition distributions, and $F_*^{-1}$ is the inverse of $F_*$ with respect to its second argument (which may be defined as in equation (38) for those $u$ that correspond to more than one $x'$). The chain's reverse transition probabilities are denoted by $\widetilde{T}^*(x^*, x^{*\prime}) = T^*(x^{*\prime}, x^*)\pi(x^{*\prime})/\pi(x^*)$, and $\widetilde{F}_*(x, x') = \int_a^{x'} \widetilde{T}^*(x, x')\, dx'$ gives the cumulative distribution functions for these reversed transitions.

Note that $y^*$ does not appear in these transition equations. It appears in a term on the right of equation (24), but this term disappears as $h$ goes to zero. The update for $y^*$ in equation (23) becomes undefined as $h$ goes to zero, as it depends on low-order parts of $u^*$ that disappear in this limit.

However, with $y^*$ eliminated, the transition on the space of $x^*$ and $u^*$ defined by equations (38) and (39) does not preserve volume. To fix this, we need to retain the part of $u^*$ that disappears in the limit as an additional part of the extended state, denoted as $v^*$, a value in $[0, 1)$. We also introduce values $t^*_0, t^*_1, t^*_2, \ldots$, which like $s^*_0, s^*_1, s^*_2, \ldots$ are independently drawn from the uniform distribution on $[0, 1)$, and which form part of the specification of the random one-to-one volume-preserving map that we can view the transitions as being. We can then define the missing part of the transition as follows:

$$y^*_{i+1} = \pi^*(x^*_{i+1})\, v^*_i \tag{40}$$

$$v^*_{i+1} = t^*_i + y^*_i/\pi^*(x^*_i) \pmod{1} \tag{41}$$

Appendix D shows that equations (38) to (41) define a map on the state space $\{(x^*, u^*, y^*, v^*) : x^* \in (a, b),\ u^* \in [0, 1),\ y^* \in [0, \pi^*(x^*)),\ v^* \in [0, 1)\}$ that is one-to-one and volume preserving, with $s^*_i$ and $t^*_i$ considered fixed. It follows from this that the transition leaves invariant the distribution in which $x^*$ has marginal probabilities given by $\pi^*$, $y^*$ given $x^*$ is uniform over $[0, \pi^*(x^*))$, and $u^*$ and $v^*$ are independent of the other parts of the state and uniformly distributed over $[0, 1)$. Note that it may be more efficient in a program to maintain $y^*/\pi(x^*)$ rather than $y^*$, since equations (40) and (41) then become simply $(y^*_{i+1}/\pi(x^*_{i+1})) = v^*_i$ and $v^*_{i+1} = t^*_i + (y^*_i/\pi(x^*_i)) \pmod{1}$, eliminating a multiply, a divide, and the need to evaluate $\pi$. In some contexts, such as parallel simulation as discussed below, it may not be necessary to maintain $y^*$ and $v^*$ at all, since they are not needed when updating $x^*$ and $u^*$.

This map can easily be adapted to a state space of all real numbers by letting $a$ go to $-\infty$ and $b$ go to $\infty$.

The map defined by equations (38) and (39) is the same as that defined by Murray and Elliott (2012) in their equations (9), (10), and (11). They do not include variables equivalent to $y^*$ and $v^*$, however, and so do not produce a volume-preserving map. For their application, and others such as parallel simulation with a single random number stream, this may not matter, but volume-preservation is needed for the importance sampling application discussed below.

MCMC is rarely used to sample one-dimensional distributions, since other methods of sampling are generally preferable, but the map described above can also be used to express an update of a single real component of a multi-dimensional state. Gibbs sampling (Gelfand and Smith, 1990) is one popular method of this sort, in which the transition updating component $j$ of a multi-dimensional state, $w$, sets it to a value drawn independently from the conditional distribution of component $j$ given the current values of other components. The map defined by equations (38) to (41) simplifies in this case, since the transitions are reversible, and ignore the previous value of component $j$.

Let $\pi(w)$ be proportional to the probability density for the multi-dimensional state $w$, and define $w \mid_j x^*$ to be state $w$ with component $j$ replaced by the value $x^*$. We can then define $\pi^*$ and $F_*$ for use in the transition from state $w_i$ as follows:

$$\pi^*(x^*) = \pi(w_i \mid_j x^*) \tag{42}$$

$$F_*(x^*) = \int_{-\infty}^{x^*} \pi^*(x^{*\prime}) \, dx^{*\prime} \bigg/ \int_{-\infty}^{\infty} \pi^*(x^{*\prime}) \, dx^{*\prime} \tag{43}$$

The update for component $j$ based on the map of equations (38) to (41) can then be expressed as

$$x_{i+1}^* = F_*^{-1}(u_i^*) \tag{44}$$

$$u_{i+1}^* = s_i^* + F_*(x_i^*) \pmod{1} \tag{45}$$

$$y_{i+1}^* = \pi(x_{i+1}^*) \, v_i^* \tag{46}$$

$$v_{i+1}^* = t_i^* + y_i^* / \pi^*(x_i^*) \pmod{1} \tag{47}$$

The multi-dimensional state after this update is $w_{i+1} = w_i \mid_j x_{i+1}^*$.

The diagram below shows such a transition (when $s_i = 0$ and $t_i = 0$):

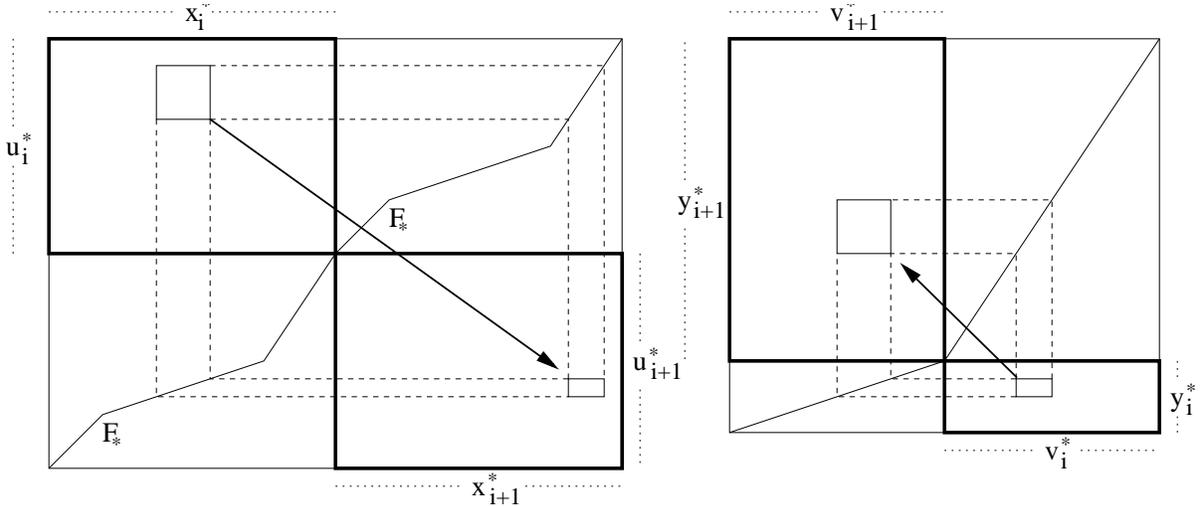

The left plot shows how, in this example, a square region of values for $x_i^*$ and $u_i^*$ maps to a smaller rectangular region for $x_{i+1}^*$ and $u_{i+1}^*$, according to the $F_*$ that is plotted. The right plot shows how a rectangular region of values for $y_i^*$ and $v_i^*$ maps to a larger square region for $y_{i+1}^*$ and $v_{i+1}^*$, according to $\pi(x_i^*)$ and $\pi(x_{i+1}^*)$, represented by the slopes of the lines shown, which in this example match the slopes of the corresponding parts of $F_*$ (as happens when $\int_{-\infty}^{\infty} \pi^*(x^{*\prime}) \, dx^{*\prime} = 1$; in general they differ by a constant factor that cancels). Note that the area on the left and the area on the right change by factors that cancel, so the transition preserves volume.

Transitions defined by equations (44) to (47) can be efficiently implemented when the conditional cumulative distribution function, $F_*$, and its inverse can be computed efficiently. In comparison, a standard implementation of Gibbs sampling will be efficient whenever it is possible to efficiently sample from this conditional distribution. For some distributions, such as the exponential and normal distributions, computing $F_*^{-1}(u)$ for a $u$ drawn from the uniform distribution on $[0, 1)$ is close to being the most efficient way of generating a random variate, in which case equation (44) has the same cost as a standard Gibbs sampling update. The cost of doing a volume-preserving Gibbs sampling update is then the extra time for equations (45) to (47), or only the time for equation (45) in a context, such as discussed below for parallel simulation, where we do not actually need to evaluate equations (46) and (47).

Metropolis updates are another common way of defining transitions for multi-dimensional states. The proposal distributions used may change only a single variable, or a subset of variables, or all variables simultaneously. If we regard such a proposal distribution as a mixture of proposal distributions $S_\delta$, as in equation (35), with the states proposed by each $S_\delta$ coming from a small set, then a volume-preserving form of the transition can be obtained in the same way as described there for a discrete state space. In particular, random-walk proposals as in equation (36) can be used, with the operation $\oplus$ being ordinary scalar or vector addition, or some other group operation such as addition modulo 1.

## 5 Application to parallel or vectorized simulation

As a first application of permutation MCMC, I will show how it can be used in parallel Markov chain simulations from multiple initial states, done with a single stream of random numbers. Using a single stream of random numbers may be faster than using separate random number streams, especially if the parallelization takes the form of vector operations, and it also avoids the problem of ensuring that multiple random streams are independent (see the discussion by Wu and Huang (2006)). However, if a single random stream is used for MCMC simulation in the standard fashion, the chains will often coalesce to the same state at some time, and thereafter track each other exactly, or they may approach each other and then move together, with states that are similar, but not identical. This would eliminate much or all of the benefit of simulating multiple chains. Such behaviour is not possible with permutation MCMC, as it would require multiple states to map to the same state, or for the mapping to be contractive, and hence not preserve volume.

To illustrate parallel simulation using permutation MCMC for a discrete state, I will use an Ising model — which has seen extensive use in statistical physics as a model of magnetization, and which is also used in image restoration (Geman and Geman, 1984).

In the Ising model, the state variables — called "spins" in the physics application — can take the values $-1$ and $+1$, and are arranged in a two-dimensional array with $r$ row and $c$ columns. The joint distribution over the $n = r \times c$ spins, $x^{(1)}, \ldots, x^{(n)}$, is defined in terms of the following "energy" function:

$$E(x) \;=\; -\sum_{(a,b)\,\in\,\mathcal{N}} x^{(a)} x^{(b)} \tag{48}$$

where $\mathcal{N}$ is a set of unordered pairs of spins that are "neighbors" — adjacent either horizontally or vertically. A spin in the leftmost column is adjacent to the spin in the rightmost column in the same row, and similarly for spins in the topmost and bottommost rows, so that all spins have four neighbors.

The probability of a state, $x$, in the Ising model is defined using this energy function as follows:
$$P(x) \;=\; \pi(x)\,/\,Z_\beta, \quad \text{with } \pi(x) \;=\; \exp(-\beta E(x)) \tag{49}$$
Here, $\beta$ is some constant, and $Z_\beta = \sum_x \pi(x)$ is the normalizing factor for the un-normalized probabilities, $\pi(x)$.

Gibbs sampling is one commonly-used MCMC method for Ising systems. Each spin is updated in some order, with the new value for the spin being randomly drawn from its conditional distribution given the values for the other spins, which is as follows:
$$P(x^{(a)} = +1 \mid x^{(b)} \text{ for } b \neq a) \;=\; \Bigg[1 \,+\, \exp\Big(-2\beta \sum_{\substack{b \text{ s.t.} \\ (a,b) \in \mathcal{N}}} x^{(b)}\Big)\Bigg]^{-1} \tag{50}$$

Since for each update there are only two possible next states, both standard Gibbs sampling and permutation Gibbs sampling using equations (22) to (24) are easily implemented. Appendix E contains an R program for implementing vectorized Gibbs sampling for the Ising model, both in the standard way, with as many random number streams as chains (all subsets of one stream in this implementation), or in the standard way, but with the same random stream for all chains, or using a single random stream with permutation MCMC. Note that the current value of the spin being updated is ignored, and that each update is reversible (though the sequential combination of such updates is not), which lead to some simplification of the general permutation MCMC transitions of equations (22) to (24).

For this demonstration, I used a 4×5 array of spins and set $\beta$ to 0.4. Figure 1 shows six Gibbs sampling simulations (distinguished by colour) done using standard MCMC (with different random numbers for each simulation), coupled MCMC (same random numbers with standard method), and permutation MCMC (same random numbers using permutation updates). All simulations were started in a state where each spin was set independently, with equal probabilities for $-1$ and $+1$. For permutation MCMC, initial values for $u^*$ and for $y^*/\pi(x)$ were drawn from the uniform distribution on $(0, 1)$.

The left plots in Figure 1 show the total energy, the right plots the magnetization (the sum of spins), after each of 250 iterations, where each iteration consists of an update of each spin in turn.

The set of standard simulations with multiple random number streams (top) is visually indistinguishable from the set of permutation MCMC simulations with a single random number stream (bottom). In contrast, the coupled chains simulated in the standard way with a single stream (middle) quickly coalesce to the same state, and thereafter provide no more information than a single chain.

I did a larger experiment with 100 parallel chains, each running for 1000 iterations. From the runs done using each method, I estimated the expectation of the energy, the expectation of the magnetization, and the expectation of the absolute value of the magnetization, along with standard errors for these estimates, computed assuming that the 100 parallel chains are independent. The results were as follows:

|                    | Energy              | Magnetization     | \|Magnetization\|  |
|--------------------|---------------------|-------------------|--------------------|
| Standard MCMC      | $-27.003 \pm 0.071$ | $+0.006 \pm 0.339$ | $14.769 \pm 0.040$ |
| Coupled MCMC       | $-27.725 \pm 0.002$ | $+4.591 \pm 0.027$ | $15.186 \pm 0.001$ |
| Permutation MCMC   | $-26.914 \pm 0.070$ | $-0.389 \pm 0.336$ | $14.720 \pm 0.040$ |
| Symmetry / Long run | $-26.944 \pm 0.020$ | $0$              | $14.746 \pm 0.012$ |

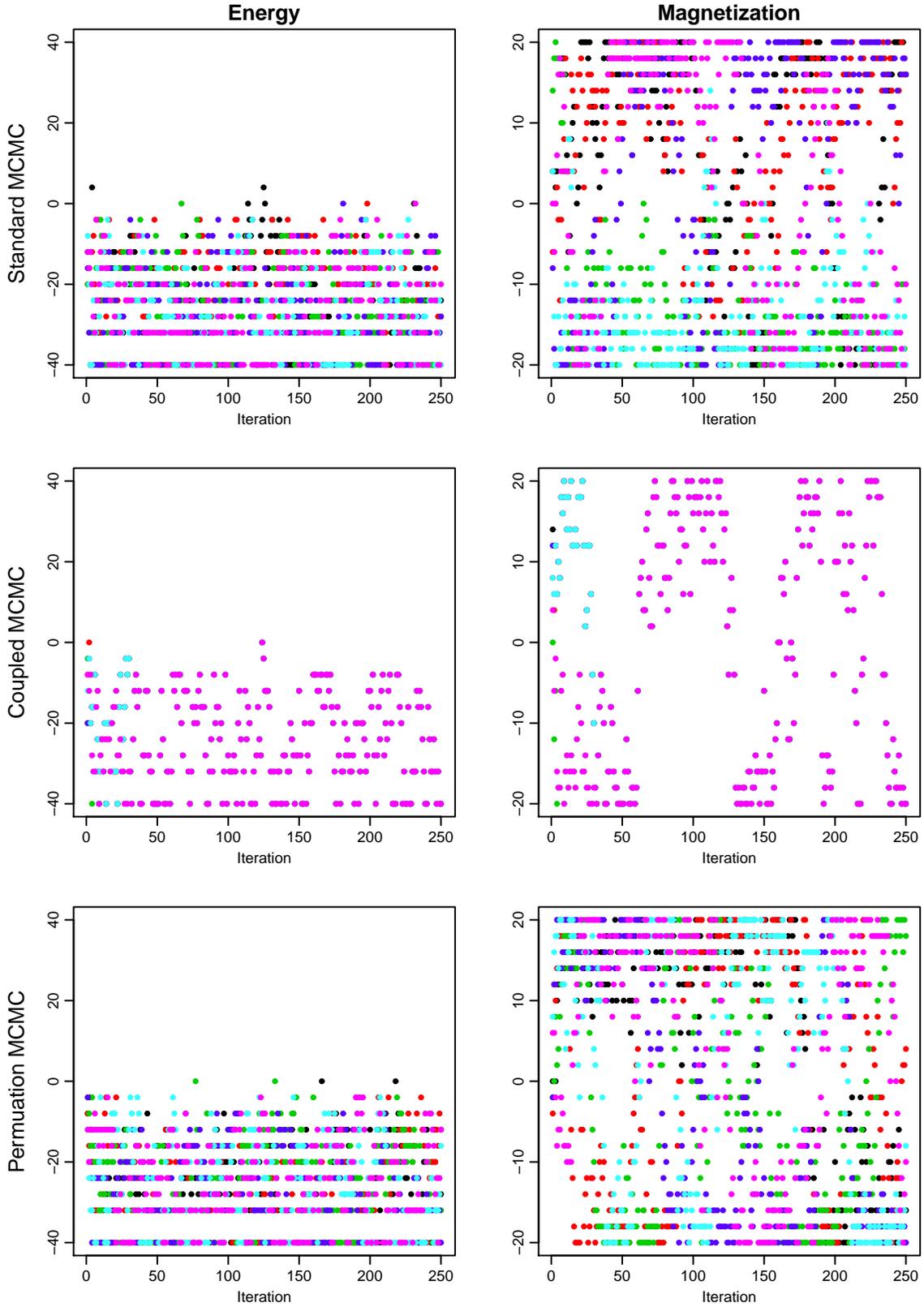

Figure 1: Six Gibbs sampling simulations (in different colours) for a $4 \times 5$ Ising model with $\beta = 0.4$, using standard MCMC, coupled MCMC, and permutation MCMC. The trace plots show energy (left) and magnetization (right) after each iteration.

The last line above gives more precise values. From symmetry, the true expectation of the magnetization is zero. For the energy and the absolute value of the magnetization, a more precise estimate of expectation was obtained by using a standard MCMC run that simulated 200 chains for 5000 iterations. Within their standard errors, the results above from standard MCMC and permutation MCMC are consistent with these more precise values (and with each other). The standard errors from standard MCMC and permutation MCMC are very close. As one would expect, the results from coupled MCMC are much less accurate, and the standard errors based on assuming the chains are independent are much too small.

I will illustrate parallel simulation using permutation MCMC on a continuous distribution using the bivariate normal with means of zero, standard deviations of one, and correlation 0.95, truncated to the interval $(-1, 2.5)$ for the first coordinate and the interval $(-1.5, 2)$ for the second coordinate. I tried both Gibbs sampling and single-variable Metropolis updates, using the program in Appendix F.

Figure 2 shows Gibbs sampling simulations for this truncated normal distribution. The permutation MCMC method for this simulation uses equations (38) and (39) only — values for $y^*$ and $v^*$ are not needed in this application, and hence neither are equations (40) and (41).

Coupled Gibbs sampling for this distribution leads to all chains approaching each other very closely within a few iterations. In contrast, the simulation done with permutation Gibbs sampling is visually indistinguishable from standard Gibbs sampling with separate random number streams.

Figure 3 shows single-variable Metropolis sampling simulations for the truncated normal distribution. In the permutation MCMC version, each coordinate update is done using equations (28) to (32), using a random-walk proposal as in equation (36), with the same $\delta$ values used for all chains, chosen randomly from the normal distribution with mean zero and standard deviation 4.

The coupled Metropolis simulation again shows a strong dependence between the chains, although unlike Gibbs sampling, the states for different chains remain far enough apart to distinguish in the plot. The permutation Metropolis simulation with a single random number stream appears visually similar to the standard Metropolis simulation with multiple streams.

In a larger experiment, I ran each method with 100 parallel chains, for 1000 iterations. The expectations of the two coordinates, and of their squares, were estimated from iterations after the first ten, and standard errors for these estimates were found on the assumption that the chains are independent. The results were as follows:

|  | 1st coord | 2nd coord | 1st squared | 2nd squared |
|---|---|---|---|---|
| Standard Gibbs sampling | $0.2274 \pm 0.0083$ | $0.2109 \pm 0.0083$ | $0.5775 \pm 0.0073$ | $0.5909 \pm 0.0067$ |
| Coupled Gibbs sampling | $0.1249 \pm 0.0002$ | $0.1011 \pm 0.0002$ | $0.4955 \pm 0.0003$ | $0.5102 \pm 0.0002$ |
| Permutation Gibbs sampling | $0.2333 \pm 0.0070$ | $0.2156 \pm 0.0072$ | $0.5874 \pm 0.0072$ | $0.6013 \pm 0.0066$ |
| Standard Metropolis | $0.2511 \pm 0.0144$ | $0.2355 \pm 0.0149$ | $0.5997 \pm 0.0125$ | $0.6144 \pm 0.0120$ |
| Coupled Metropolis | $0.4259 \pm 0.0074$ | $0.4016 \pm 0.0076$ | $0.7105 \pm 0.0092$ | $0.6883 \pm 0.0089$ |
| Permutation Metropolis | $0.2658 \pm 0.0162$ | $0.2468 \pm 0.0164$ | $0.6243 \pm 0.0155$ | $0.6354 \pm 0.0143$ |
| Long simulation run | $0.2329 \pm 0.0012$ | $0.2162 \pm 0.0012$ | $0.5821 \pm 0.0011$ | $0.5962 \pm 0.0010$ |

The more precise estimates in the last line above are from a standard Gibbs sampling run that simulated 200 chains for 20000 iterations.

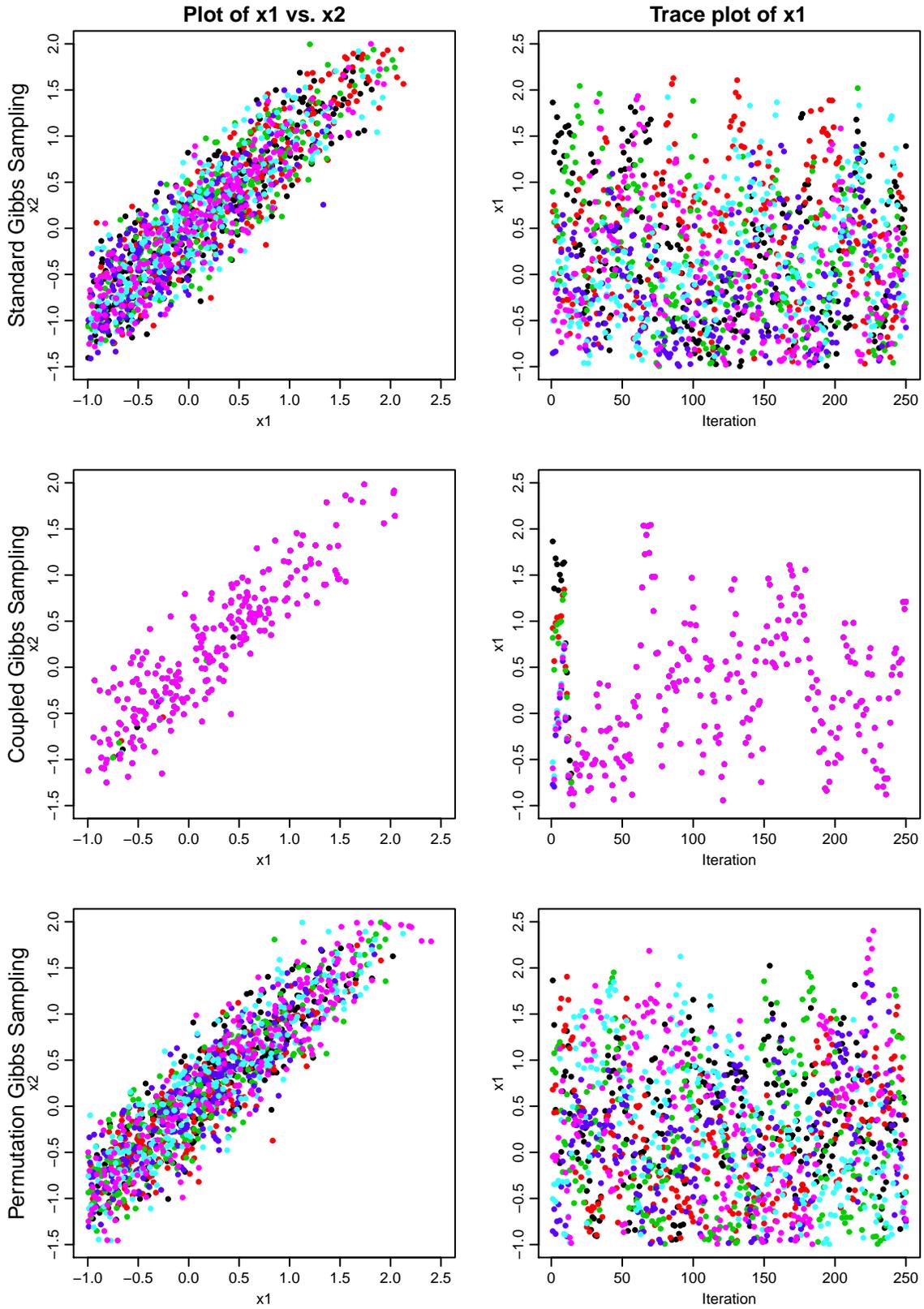

Figure 2: Six Gibbs sampling simulations for a truncated bivariate normal distribution, using standard MCMC, coupled MCMC, and permutation MCMC.

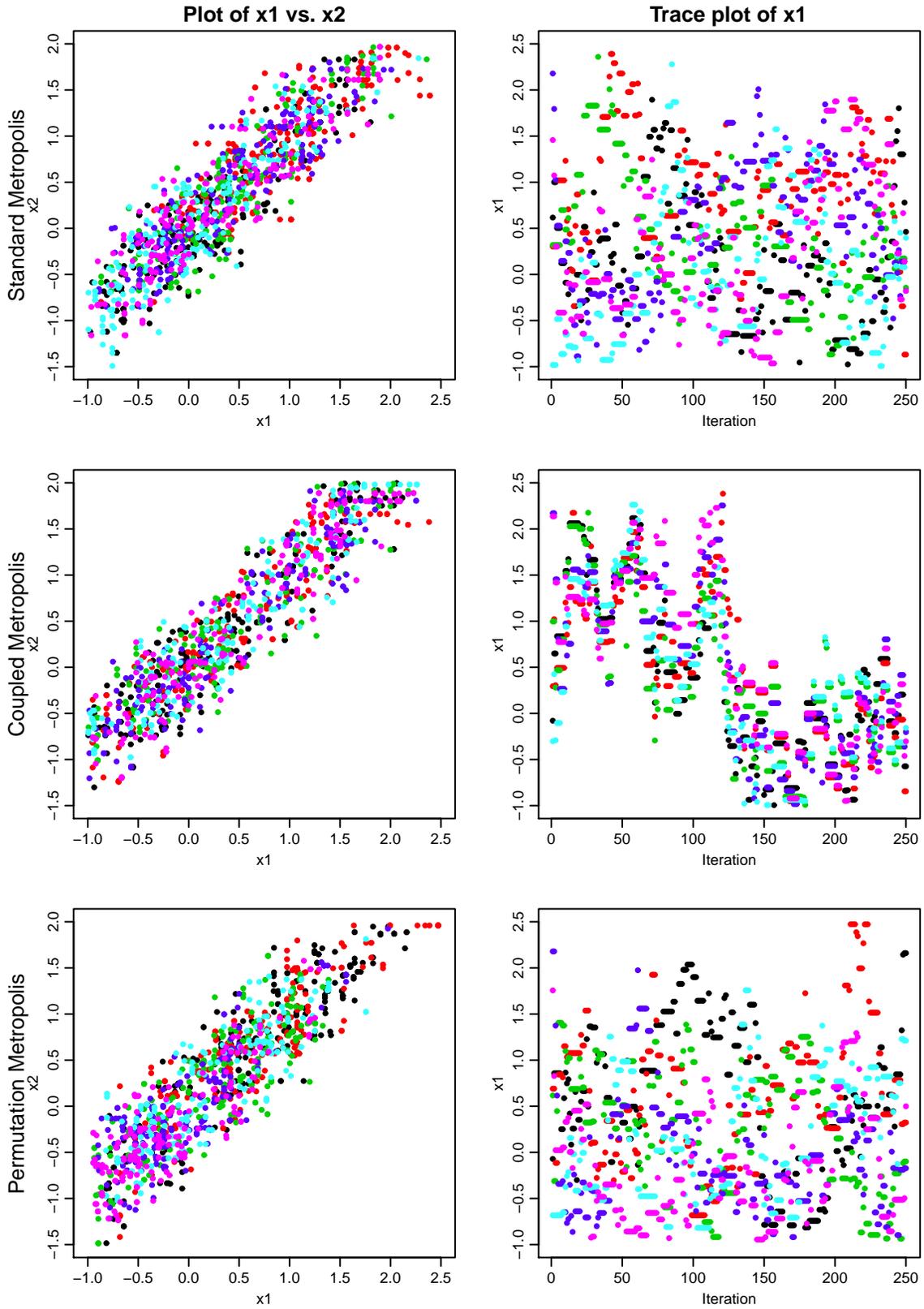

Figure 3: Six single-variable Metropolis simulations for a truncated bivariate normal distribution, using standard MCMC, coupled MCMC, and permutation MCMC.

From this table, we see that, as for the Ising model, coupled MCMC does not work well for a parallel simulation with a single stream, but the accuracy of permutation Gibbs sampling and permutation Metropolis with a single random stream is close to that of standard simulation with multiple random streams. (In the table, there appears to be a slight advantage for permutation Gibbs sampling, and a slight disadvantage for permutation Metropolis, but neither is seen consistently in replications with different random seeds.)

Will parallel simulation using permutation MCMC with a single random number stream always behave similarly to standard MCMC with multiple streams? For simplicity, consider the case of just two chains. With standard MCMC with two random streams, the states of these chains will be independent (assuming independent initial states). The map defined by permutation MCMC on its extended state space (consisting of $x$ or $x^*$, $u$ or $u^*$, and possible $y$ or $y^*$, etc.) will be a permutation (or a volume-preserving map). The map on the joint space for two chains will also be a permutation (or will be volume-preserving), and hence will leave the uniform distribution on the pair of (extended) states for the two chains invariant. In this uniform joint distribution, the states of the two chains are independent.

However, this does not imply that parallel simulation with permutation MCMC will produce chains with independent states, because the map on the joint space produced by permutation MCMC may not be ergodic. Indeed, if the extended state space is finite (as in Section 2), it cannot be ergodic — if the two chains are in the same state, they will remain in the same state, so there are at least two disconnected regions of the joint state space that both have non-zero probability.

This particular non-ergodicity will actually make permutation MCMC perform better than standard MCMC, provided the initial states of the chains are chosen to be different, since it will induce negative correlations that improve the accuracy of estimates (though the effect is probably negligible in real problems with huge state spaces). However, one may wonder whether there might sometimes be other non-ergodic aspects of the permutation MCMC map as applied to multiple states, which might lead to worse performance compared to standard parallel MCMC. Of course, as long as the MCMC method is ergodic for one chain, the results found using permutation MCMC will be valid — only the degree of improvement in accuracy from using multiple chains is at issue.

## 6 Application to importance sampling

Another application of permutation MCMC is to improving a deficient, but not too deficient, importance sampling distribution.

Recall that the expectation of some function $a(x)$ with respect to some distribution of interest, with probability density (or mass) function proportional to $\pi(x)$, can be estimated from points $x_1, \ldots, x_N$ drawn independently from some importance sampling distribution, whose probability density (or mass) function is proportional to $\rho(x)$, as follows:

$$E_\pi[a(x)] \approx \frac{\sum_{i=1}^{N} a(x_i)\pi(x_i)/\rho(x_i)}{\sum_{i=1}^{N} \pi(x_i)/\rho(x_i)} \qquad (51)$$

Methods for assessing the accuracy of such estimates are discussed by Geweke (1989) and Neal (2001, Section 3).

Suppose that we have found some tractable importance sampling distribution — for which it is feasible to sample independent points, and to compute $\rho(x)$ for these sampled points — but that this importance sampling distribution gives very low probability to some region that has non-negligible probability under the distribution of interest. Such an importance sampling distribution will not work well, since it provides little or no information about this region, even though it cannot be ignored.

I will demonstrate here that permutation MCMC can be used to improve such a deficient importance sampling distribution, as long as there is substantial overlap between the original importance sampling distribution and the distribution of interest (that is, their total variation distance is not close to one).

The improved importance sampling distribution is defined as the mixture of distributions obtained by sampling from the original importance sampling distribution and then applying a permutation MCMC map for some number of iterations between 0 to $M$, chosen randomly, here with equal probabilities (though unequal probabilities could be used). Since the permutation map is on an extended state space, the value for $x$ or $x^*$ sampled from the importance sampling distribution must be supplemented with a value for $u$ or $u^*$, plus for non-uniform distributions, a value for $y$ or $y^*$, and for continuous transitions, a value for $v^*$. These additional parts of the extended state are drawn uniformly from their allowed range, which is $[0, 1)$ except for $y$ or $y^*$, whose range is $[0, \pi(x))$ or $[0, \pi(x^*))$. These additional variables are drawn independently of each other and of $x$ or $x^*$ (except for the dependence of the range of $y$ or $y^*$ on $x$ or $x^*$).

Before drawing points from the improved importance sampling distribution, we must fix particular permutation MCMC maps, for up to $M$ iterations, by sampling $s_0, s_1, \ldots, s_{M-1}$, as well as $t_0, t_1, \ldots, t_{M-1}$ and $\delta_0, \delta_1, \ldots, \delta_{M-1}$ if required. We can then sample points as follows:

1) Sample $k$ uniformly from $\{0, \ldots, M\}$.

2) Sample $x_k$ or $x_k^*$ according to $\rho$.

3) Sample other variables in the extended state (eg, $u_k^*$) uniformly from their allowed range.

4) Simulate $M - k$ permutation MCMC transitions, using $s_k, \ldots, s_{M-1}$ (and also $t_k, \ldots, t_{M-1}$ and $\delta_k, \ldots, \delta_{M-1}$ if required), producing extended states indexed by $k+1, \ldots, M$. (If $k = M$, no transitions are simulated.)

5) Let the extended state indexed by $M$ be the point drawn from the improved importance sampling distribution.

Supposing that it is feasible to compute $\rho(x)$, the probability density (or mass) of the point sampled above under the improved improved importance sampling distribution can also be computed, with the following additional steps:

6) Do a *reverse* permutation MCMC simulation for $k$ iterations, starting with the originally sampled $x_k$ or $x_k^*$, along with the other variables in the extended state indexed by $k$, using $s_{k-1}, \ldots, s_0$ (and also $t_{k-1}, \ldots, t_0$ and $\delta_{k-1}, \ldots, \delta_0$ if required), producing extended states indexed by $k-1, \ldots, 0$. (If $k = 0$, no reverse transitions are simulated.)

7) For $j = 0, \ldots, M$, compute $\rho(x_j)$ or $\rho(x_j^*)$, and then, for a uniform distribution, compute

$$\ddot{\rho} \;=\; \frac{1}{M+1} \sum_{j=0}^{M} \rho(x_j) \qquad (52)$$

and for a non-uniform distribution, compute

$$\ddot{\rho} = \frac{1}{M+1} \sum_{j=0}^{M} \frac{\rho(x_j)}{\pi(x_j)} \tag{53}$$

or similarly for a continuous state, $x^*$.

The extended states indexed by $0, \ldots, M$ are the possible start states that map to the state indexed by $M$, when between $M$ and $0$ permutation MCMC iterations are done. Since each map is a permutation, or preserves volume, the probability of obtaining the extended state indexed by $M$ with some number of iterations is simply the probability of sampling the $x_j$ in the corresponding start state from the original importance sampling distribution, times the probability of sampling the additional variables in the extended state.

The total (possibly unnormalized) probability mass or density for obtaining the sampled point from the improved importance sampling distribution is $\ddot{\rho}$, which simply adds the probabilities of obtaining the point from all possible start states. The division by $\pi(x_j)$ in equation (53) results from the range of $y_j^*$ being $[0, \pi(x_j))$, so that the density of $y_j^*$ is $1/\pi(x_j)$.

If we perform the above procedure $N$ times, we will obtain $N$ points from step (5), which we can label $x_i$ for $i = 1, \ldots, N$, along with corresponding quantities $\ddot{\rho}_i$ from step (7). The desired distribution on the extended state is uniform ($\pi$ being accounted for in the range of $y^*$), so the improved importance sampling estimate for the expectation of $a(x)$ is

$$E_\pi[a(x)] \approx \frac{\sum_{i=1}^{N} a(x_i)/\ddot{\rho}_i}{\sum_{i=1}^{N} 1/\ddot{\rho}_i} \tag{54}$$

When $M = 0$, this reduces to the original importance sampling estimate of equation (51).

I will demonstrate how this method can sometimes improve an importance sampling distribution using a two-dimensional test distribution, in which the first coordinate, $x^{(1)}$, is normally-distributed with mean zero and standard deviation one, and given a value for the first coordinate, the second coordinate, $x^{(2)}$, is normally-distributed with mean $[x^{(1)}]^2 - 1$ and standard deviation one. I will consider using the method to improve on four importance sampling distributions, all of which are bivariate normal with zero correlation and equal standard deviations for the two coordinates, but which differ in the means for the two coordinates and in the standard deviation of the coordinates. Figure 4 shows samples of 500 points from the four importance sampling distributions (in red), each shown along with a sample of points from the test distribution (in black).

For MCMC sampling from this test distribution, I use random-walk Metropolis transitions that update both coordinates simultaneously, with the proposal offset drawn from the bivariate normal distribution with zero correlation, zero mean, and standard deviation of 4 for both coordinates. The permutation MCMC implementation of this transition is done by fixing $\delta_0, \delta_1, \ldots, \delta_{M-1}$ to values drawn from this proposal offset distribution, and using them as described at the end of Section 3. The program for this permutation MCMC method and the importance sampling procedure above that uses it is shown in Appendix G.

Figure 5 shows distributions obtained by drawing points from two of the importance sampling distributions and then applying between zero and 10 iterations of this permutation MCMC map. The top plots are for an importance sampling distribution (leftmost in Figure 4) that is quite concentrated (standard deviation 0.3) and located in an area of the test distribution of fairly high probability. The bottom plots are for an importance sampling distribution (rightmost in Figure 4) with the same standard deviation, but located in an area of very low probability under the test distribution. The four plots in each row show the importance sampling distributions obtained with four different random selections of $s_0, \ldots, s_9$ and $\delta_0, \ldots, \delta_9$.

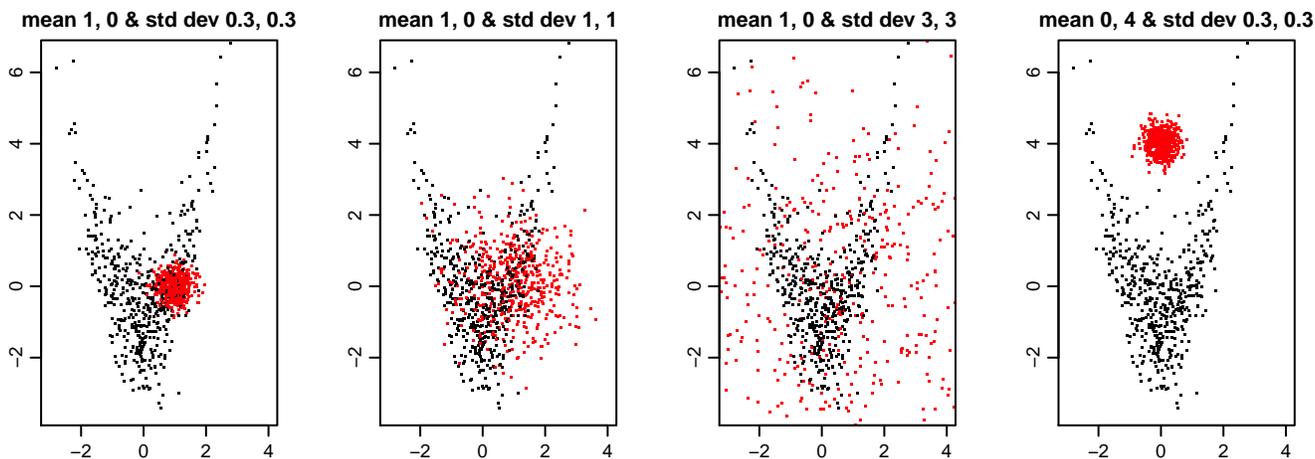

Figure 4: Four initial importance sampling distributions for the test distribution. Each plot shows 500 black dots from the test distribution and 500 red dots from initial independent normal importance sampling distributions with means and standard deviations as shown.

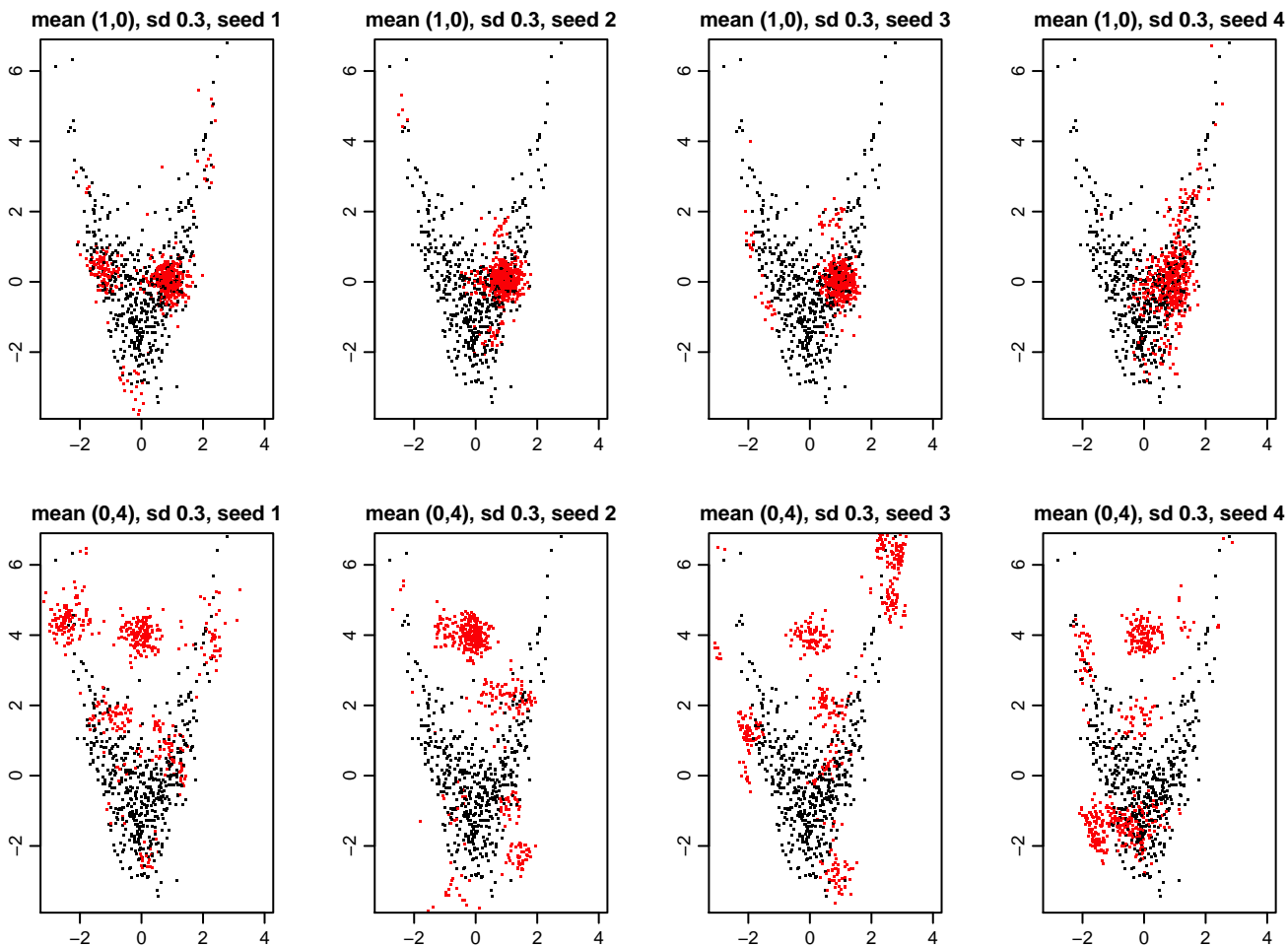

Figure 5: Importance sampling distributions obtained by applying from 0 to 10 permutation MCMC iterations starting from the leftmost and rightmost importance sampling distributions in Figure 4. The four plots in each row are for different random selections of $s_0, \ldots, s_9$ and $\delta_0, \ldots, \delta_9$.

As can be seen in these plots, up to 10 iterations is not enough for the points sampled from these overly-concentrated importance sampling distributions to be transformed to have a distribution that is close to the test distribution. As the number of iterations is increased, the distribution for $x^{(1)}$ and $x^{(2)}$ will approach the test distribution, but this alone is not enough to ensure that importance sampling will work well — the distribution of the entire extended state must be suitable. One way to visualize the adequacy of the importance sampling is to imagine drawing values for $x^{(1)}$ and $x^{(2)}$ from the test distribution, along with a value for $u^*$ uniformly from $[0, 1)$ and a value for $y^*$ uniformly from $[0, \pi(x^{(1)}, x^{(2)}))$, and then applying the reverse permutation MCMC map for $M$ iterations from this state. For the improved importance sampler to be adequate, it must almost always be the case that at least one of the states from the reverse simulation has values for $x^{(1)}$ and $x^{(2)}$ that have reasonably high probability under the original importance sampling distribution.

Figure 6 shows results on the test distribution using the four original importance sampling distributions of Figure 4, focusing on estimating the expectation of $[x^{(2)}]^2$, whose true value is exactly 3. For each original distribution, four values of $M$ were tried, with $M = 0$ being equivalent to just using the original importance sampling distribution. For each value of $M$, results using three random seeds are shown, which differ both in the points sampled from the original importance sampling distribution and in the random choice of $s_0, \ldots, s_{M-1}$ and $\delta_0, \ldots, \delta_{M-1}$. The estimates were based on $N = 2000$ points. Standard errors for the estimates were calculated with the procedure at the end of Appendix G.

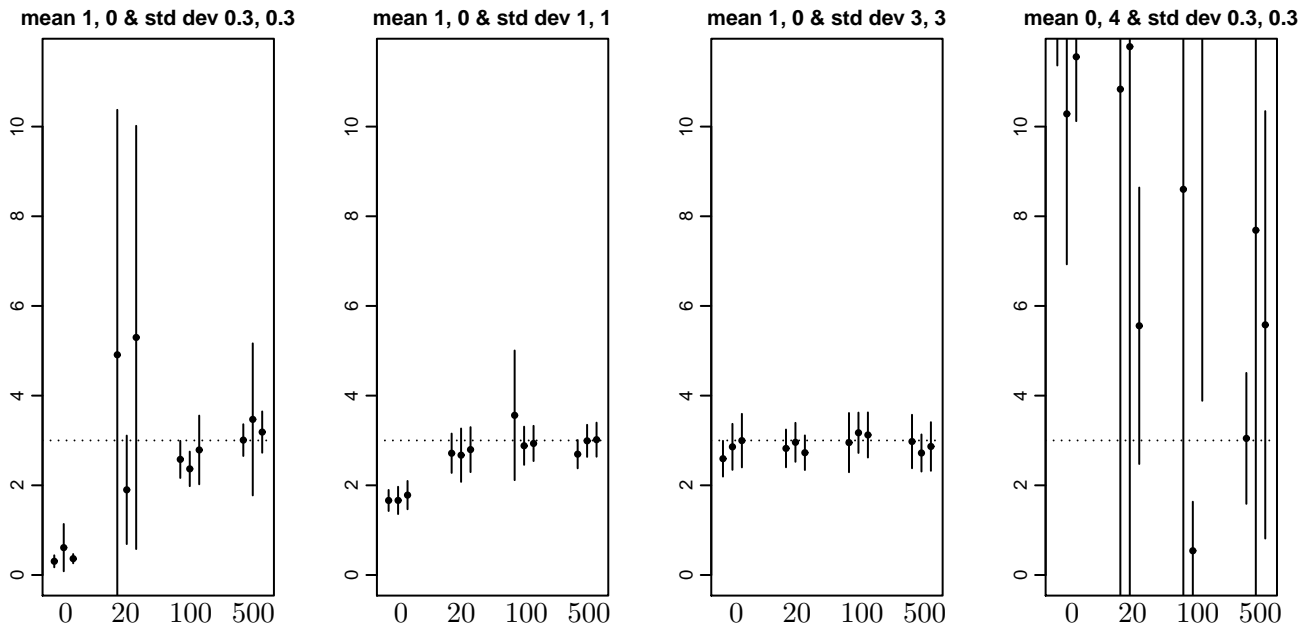

Figure 6: Estimates for the expectation of $[x^{(2)}]^2$ from improved importance sampling using the four original importance sampling distributions shown in Figure 4, and with $M$ set to 0 (equivalent to the original distribution), 20, 100, and 500. Estimates with $N = 2000$ are shown by dots, with vertical lines extending up and down by twice the standard error. For each value of $M$, estimates from three random settings of $s_0, \ldots, s_9$ and $\delta_0, \ldots, \delta_9$ are shown. The true expection of 3 is marked with a dotted line.

From the results with $M = 0$, we see that of the original importance sampling distributions, only the third, with standard deviations of 3, produces adequate results. The more concentrated importance sampling distributions give very low probability to some high-probability regions of the test distribution, with the result that the estimate obtained is far from the truth, and moreover the estimated standard error is much too small.

The first and second importance sampling distributions are improved by applying permutation MCMC. For the first importance sampling distribution, using $M = 20$ produces estimates that are rather inaccurate, but at least the standard errors properly reflect this. Good estimates are obtained when $M = 100$ or $M = 500$. For the second importance sampling distribution, even $M = 20$ gives reasonably good estimates, and realistic standard errors, and the estimates improve (with some variation) with larger $M$. The improvement seen is as expected based on the discussion above, since 20 or more Metropolis iterations started at a point drawn randomly from the test distribution are indeed likely to produce at least one state from the region with high probability under either of these two original importance sampling distributions, since they have substantial overlap with the test distribution.

In contrast, for the fourth importance sampling distribution, only a slight improvement is seen from using permutation MCMC. Even with $M = 500$, the estimates are very inaccurate, though perhaps the standard errors are more realistic than for $M = 0$. This is also as expected given the discussion above — the fourth importance sampling distribution is concentrated almost entirely in a low-probability area of the test distribution, so even 500 Metropolis iterations from a point randomly sampled from the test distribution are likely to all remain in regions with low probability under the original importance sampling distribution.

Permutation MCMC also produces only a slight improvement for the the third importance sampling distribution. This distribution is diffuse enough to cover the entire test distribution. Such a simple diffuse importance sampling distribution will usually be ineffective for a high-dimensional problem, because too many points drawn from it will land in regions of very low probability under the distribution of interest. But for this two-dimensional test problem the effective sample size is reduced by only a factor about 5.5, so acceptable results are obtained. We might hope, nevertheless, that applying permutation MCMC would improve performance, but even for $M = 500$, the inefficiency factor is reduced only to about 3.0.

This might seem puzzling, since a point drawn from this diffuse importance sampling distribution is likely to map to a point that has high probability density under the test distribution within a fairly small number of Metropolis updates. However, points derived from original points with low density under the test distribution have very *high* density under the improved importance sampling distribution, because the value of $y^*$ for such points is drawn from the narrow range $[0, \pi(x^{(1)}, x^{(2)}))$. Such points therefore receive a very low weight in the estimate of equation (54), reducing the effective sample size. All sampled points will receive nearly equal weight only when $M$ is so large that $M$ reverse transitions started from a point drawn randomly drawn from the test distribution are likely to contain several points with low density under the test distribution, comparable to the low density of many points drawn from the original importance sampling distribution.

This method for improving importance samplers using permutation MCMC could itself be improved in several respects. First, once random values for $s_0, \ldots, s_{M-1}$ and $\delta_0, \ldots, \delta_{M-1}$ have been chosen, $N$ points from the improved importance sampling distribution could be simulated in parallel. This is most easily done if the random choice of $k$ from $\{0, \ldots, M\}$ is replaced by stratified sampling (as is desirable in any case), so that exactly $N/(M+1)$ points are generated with each value of $k$. The simulations for each value of $k$ could then be done with the vectorized permutation MCMC procedure.

Secondly, the procedure described above actually produces results from *two* improved importance sampling distributions, only one of which is used above. The second distribution would use the state indexed by 0 after step (7) of the procedure above. More generally, several importance sampling distributions that use overlapping sequences $s_0, \ldots$ and $\delta_0, \ldots$ could be defined, and points drawn from all these distributions could be obtained with much of the computation being in common. The results from all these importance sampling distributions could then be combined.

Even with these improvements, the method described above will be restricted to improving importance sampling distributions that already have substantial overlap with the desired distribution. For complex, high-dimensional problems, finding such a distribution may be difficult. However, I expect that permutation MCMC can be used in an annealing framework, as I have previously developed for Hamiltonian importance sampling (Neal, 2005), and applied to a wide range of complex distributions, including those with multiple isolated modes. As for any importance sampling procedure, it would also be possible to estimate the ratio of the normalizing constants for the importance sampling distribution and the distribution of interest. These ratios are of great interest in both statistical physics and Bayesian inference.

# 7 The role of randomness in MCMC simulation

The initial work on Markov chain Monte Carlo by Metropolis, *et al* (1953), who used it to sample from the "canonical" distribution for a system of molecules, was soon followed by work on an alternative deterministic approach to molecular simulation (Alder and Wainwright, 1959). In the context of statistical physics, these stochastic and deterministic approaches to simulation become equivalent in the limit as the size of the system increases. It is also possible to extend the state space of a deterministic dynamical simulation so that even for small systems it visits states according to the same distribution that a stochastic simulation would (Nosé, 1984).

Permutation MCMC provides another way of reducing, or even eliminating, the role of randomness in MCMC simulation.

In a standard MCMC simulation, the initial states of the chains may be chosen randomly (though usually not from the desired distribution), or by using some non-random heuristic (such as starting at the mode), or more-or-less arbitrarily. Subsequent transitions are always random (in practice, pseudo-random) rather than deterministic. As long as the transitions are ergodic, the choice of initial states affects only the number of iterations needed to reach a good approximation of the equilibrium distribution, not the asymptotic results from averaging functions of state over subsequent iterations of the chains.

With permutation MCMC, *any* choice of $s_0, s_1, \ldots$ (along with any other variables that define the permutation or volume-preserving map) will leave the desired distribution invariant — as has also been noted by Murray and Elliott (2012) for their related MCMC method. It follows that if a set of initial states are drawn from the desired distribution (which will be uniform, if we extend the state as in this paper), estimates based on the sequences of states found by applying permutation MCMC from these initial states will be unbiased even if $s_0, s_1, \ldots$ are non-random.

In most applications, however, we are not able to choose initial states from the desired distribution. If we use some other initial state distribution, correct results will be obtained by averaging over the sequence of states obtained using permutation MCMC only if the particular sequence $s_0, s_1, \ldots$ that we use has a suitable ergodic property. Note that if $s_0, s_1, \ldots$ is considered non-random, it will *not* be the case that the distribution of the state at time $t$ will converge to the desired state as $t$ goes to infinity — these state distributions will remain just as non-uniform as the initial state distribution. Instead, we can hope that time averages of functions of state will converge to their correct expectations. Characterizing when this will occur is an interesting topic for further research, which may relate to work on quasi-Monte Carlo methods for MCMC (eg, Chen, Dick, and Owen, 2011).

Some empirical insight into this question can be obtained using the permutation MCMC procedures for the Ising model and the truncated normal distribution that were used in Section 5.

Here are the results of 100 chains, run for 1000 iterations each, using permutation Gibbs sampling for the Ising model:

|  | Energy | Magnetization | \|Magnetization\| |
|---|---|---|---|
| With $s$ set randomly | $-26.914 \pm 0.070$ | $-0.389 \pm 0.336$ | $14.720 \pm 0.040$ |
| With $s$ repeating 0.2, 0.6 | $-24.201 \pm 0.041$ | $+0.127 \pm 0.249$ | $13.633 \pm 0.027$ |
| With $s$ repeating 0.3 | $-25.512 \pm 0.048$ | $-0.025 \pm 0.268$ | $14.178 \pm 0.027$ |
| With $s$ repeating 0.213, 0.631 | $-26.876 \pm 0.061$ | $+0.055 \pm 0.303$ | $14.705 \pm 0.035$ |
| With $s$ repeating 0.292 | $-26.893 \pm 0.054$ | $-0.266 \pm 0.271$ | $14.721 \pm 0.032$ |
| With $s$ repeating 0.237 | $-26.877 \pm 0.050$ | $-0.069 \pm 0.270$ | $14.733 \pm 0.029$ |
| Symmetry / Long run | $-26.944 \pm 0.020$ | 0 | $14.746 \pm 0.012$ |

The first line above is the same as was reported for permutation MCMC in Section 5. The other lines are for runs in which $s_0, s_1, \ldots$ are set deterministically, by repeating a pair of values, or a single value. We see that incorrect results are obtained when repeating the pair 0.2, 0.6, or the single value 0.3, but results are consistent with the answer from symmetry or a long run when repeating the pair 0.213, 0.631, or the single values 0.292 or 0.237. This difference is perhaps related to 0.2, 0.3, and 0.6 being rational with small numerator and denominator, but further research is needed to understand these results.

Figure 7 shows six runs of permutation Gibbs sampling on the truncated normal distribution used in Section 5. The top plot is for permutation MCMC with $s_0, s_1, \ldots$ set randomly, the same as the bottom plot of Figure 2. The middle plot is for $s_0, s_1, \ldots$ alternately set to 0.231 and 0.452. In the bottom plot, all the $s_i$ are set to 0.211. Visually, the two deterministic sequences appear to produce correct results. Figure 8 shows more results with all $s_j$ set to a single value — to zero for the top plot, to 0.017 for the middle plot, and to 0.211 for the bottom plot (as for Figure 7). Setting all $s_j$ to zero produces interesting results, but it clearly does not result in the correct distribution. Setting all $s_j$ to 0.017 does appear to produce correct results.

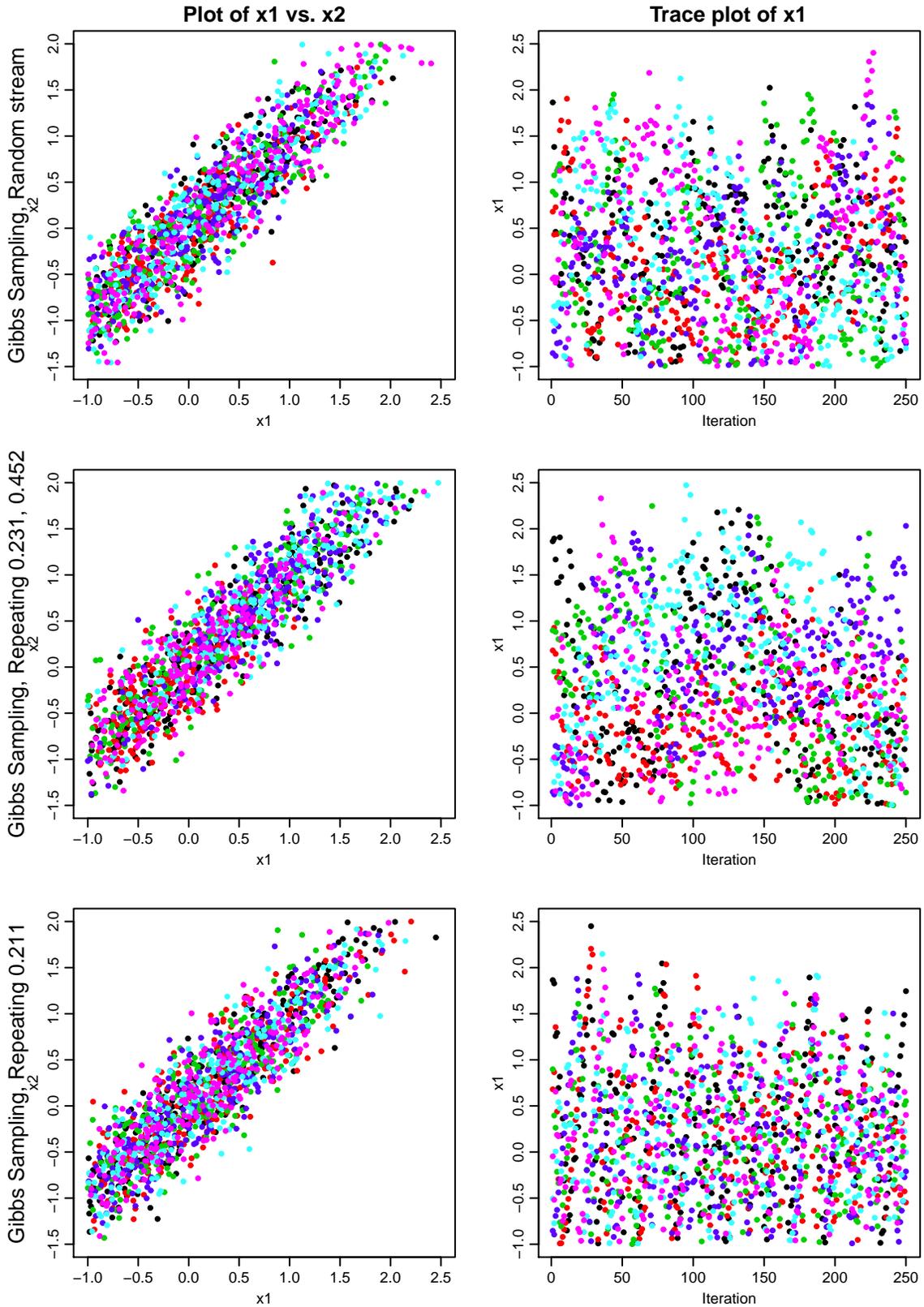

Figure 7: Six permutation Gibbs sampling simulations for a truncated bivariate normal distribution, with $s$ set randomly (as in the bottom plot of Figure 2), set to alternating values of 0.231 and 0.452, and set always to 0.211.

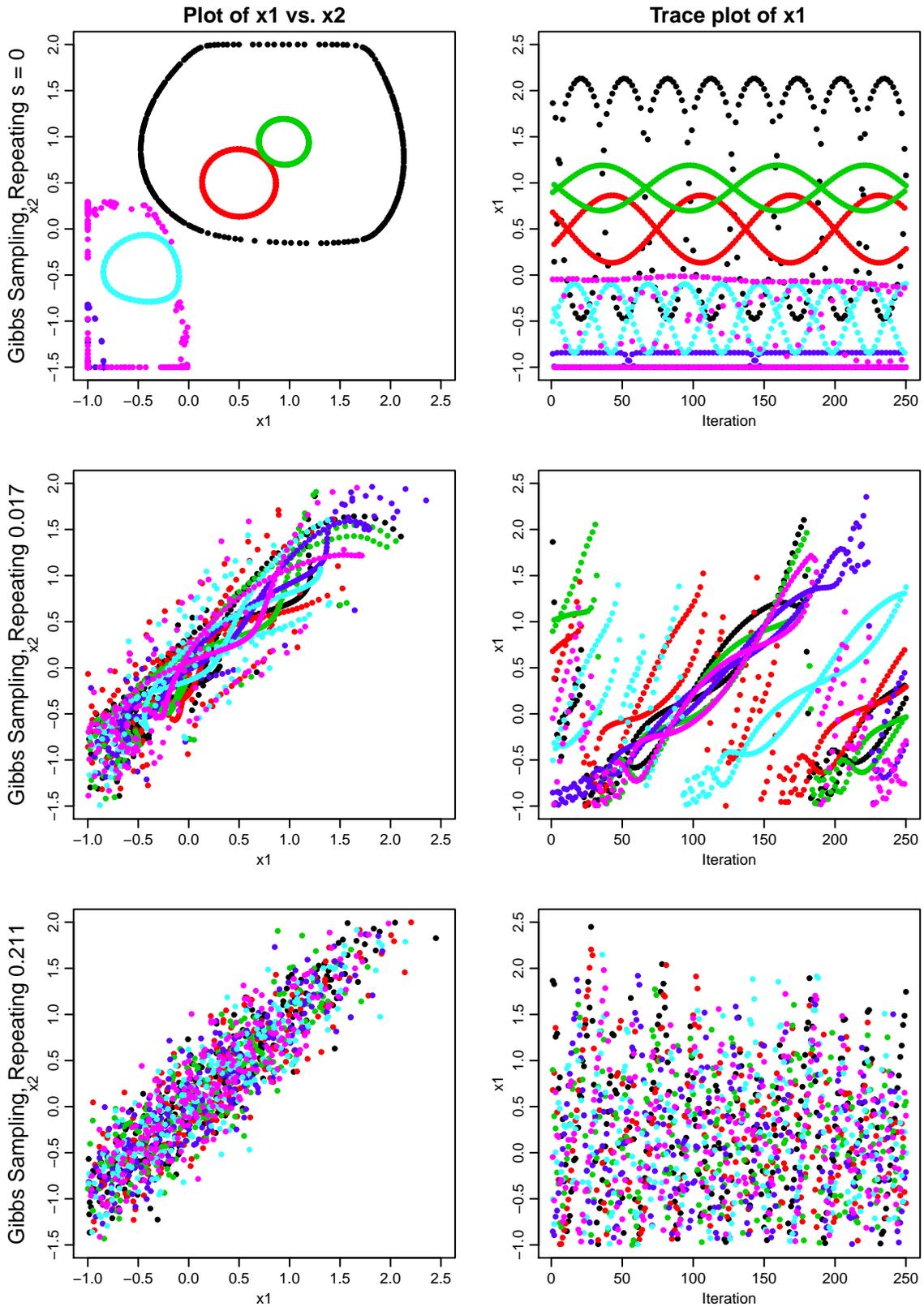

Figure 8: Six permutation Gibbs sampling simulations for a truncated bivariate normal distribution, with $s$ set always to 0, to 0.017, and to 0.211 (as in the bottom plot of Figure 7).

To confirm these visual impressions, I did 100 runs of 1000 iterations with these deterministic settings for $s_0, s_1, \ldots$. The estimates obtained, along with their standard errors, are shown below:

|  | 1st coord | 2nd coord | 1st squared | 2nd squared |
|---|---|---|---|---|
| With $s$ set randomly | $0.2333 \pm 0.0070$ | $0.2156 \pm 0.0072$ | $0.5874 \pm 0.0072$ | $0.6013 \pm 0.0066$ |
| With $s$ repeating 0.231, 0.452 | $0.2408 \pm 0.0119$ | $0.2245 \pm 0.0122$ | $0.5792 \pm 0.0094$ | $0.5944 \pm 0.0090$ |
| With $s$ always 0.211 | $0.2388 \pm 0.0023$ | $0.2216 \pm 0.0023$ | $0.5893 \pm 0.0046$ | $0.6018 \pm 0.0045$ |
| With $s$ always 0.017 | $0.2267 \pm 0.0055$ | $0.2087 \pm 0.0052$ | $0.5969 \pm 0.0112$ | $0.6110 \pm 0.0109$ |
| With $s$ always 0 | $0.4431 \pm 0.0742$ | $0.4001 \pm 0.0770$ | $1.0609 \pm 0.0974$ | $1.1092 \pm 0.0828$ |
| Long simulation run | $0.2329 \pm 0.0012$ | $0.2162 \pm 0.0012$ | $0.5821 \pm 0.0011$ | $0.5962 \pm 0.0010$ |

All the runs except the one with $s$ always set to zero produce results that are consistent with the long simulation run. However, the runs vary in efficiency, as seen from the standard errors of the estimates. Estimates with $s$ repeating 0.231, 0.452 are less efficient than when $s$ is set randomly. When $s$ is fixed at 0.211, however, the standard errors for the two coordinates are a factor of three smaller (corresponding to a factor of nine better efficiency), and the standard errors for the squares of the coordinates are a factor of 1.5 smaller (a factor of 2.25 more efficient). When $s$ is fixed at 0.017, the standard errors for the coordinates are smaller than for random $s$, but the standard errors for the squares of the coordinates are larger.

The source of the greater efficiency when $s$ is fixed at 0.017 or 0.211 can be seen from an examination of the middle and bottom plots in Figure 8. In the middle plot, for $s = 0.017$, the two coordinates move upward in a slow but consistent fashion for many iterations, and then quickly move downward, before resuming their upward motion. The same behaviour is seen in the bottom plot, for $s = 0.211$, except that the upward motion is quicker, making the pattern less obvious. Such consistent upward motion moves around the distribution more efficiently than the random walk motion characteristic of standard Gibbs sampling — $n$ steps that are taken consistently in one direction lead to a point about $n$ steps away from the start point, whereas $n$ steps taken in a random walk lead to point that is likely only about $\sqrt{n}$ steps away. In this respect the behaviour with $s$ fixed to a small value resembles that of "overrelaxation" methods (Adler, 1981; Neal, 1998), though the dynamics of overrelaxation is quite different.

Exploring permutation MCMC methods with non-random values for $s_0, s_1, \ldots$ (and $\delta_0, \delta_1, \ldots$ where needed) seems to be a promising avenue to finding more efficient MCMC methods, but better theory regarding when such non-random methods will be ergodic would be helpful. However, as noted by Murray and Elliott (2012), one can ensure ergodicity (assuming the chain is ergodic when simulated in the standard manner) by combining non-random selections of $s_j$ with occasional sequences of randomly chosen $s_j$ that are long enough to move anywhere in the state space.

## Acknowledgements

This research was supported by Natural Sciences and Engineering Research Council of Canada. The author holds a Canada Research Chair in Statistics and Machine Learning.

# Appendix A: Proof that the map for a simulation of a uniform discrete distribution is a permutation

To show that the map $(x_i, u_i) \to (x_{i+1}, u_{i+1})$ defined by equations (2) and (3) is a permutation, it suffices to show that the following map is its inverse:

$$x_i^\dagger = \max\left\{x' : Q \sum_{x=0}^{x'-1} \widetilde{T}(x_{i+1}, x) \leq u_{i+1} - s_i \pmod{Q}\right\} \tag{55}$$

$$u_i^\dagger = (u_{i+1} - s_i) - Q \sum_{x=0}^{x_i^\dagger - 1} \widetilde{T}(x_{i+1}, x) + Q \sum_{x=0}^{x_{i+1}-1} T(x_i^\dagger, x) \pmod{Q} \tag{56}$$

That is, we need to show that equations (2), (3), (55), and (56) imply that $x_i^\dagger = x_i$ and $u_i^\dagger = u_i$. Equation (2) implies that

$$0 \leq u_i - Q \sum_{x=0}^{x_{i+1}-1} T(x_i, x) < Q T(x_i, x_{i+1}) = Q \widetilde{T}(x_{i+1}, x_i) \tag{57}$$

and hence

$$0 \leq Q \sum_{x=0}^{x_i-1} \widetilde{T}(x_{i+1}, x) \leq u_i - Q \sum_{x=0}^{x_{i+1}-1} T(x_i, x) + Q \sum_{x=0}^{x_i-1} \widetilde{T}(x_{i+1}, x) < Q \sum_{x=0}^{x_i} \widetilde{T}(x_{i+1}, x) \leq Q \tag{58}$$

It follows from this and equation (3) that

$$u_{i+1} - s_i \pmod{Q} = u_i - Q \sum_{x=0}^{x_{i+1}-1} T(x_i, x) + Q \sum_{x=0}^{x_i-1} \widetilde{T}(x_{i+1}, x) \tag{59}$$

Hence we can rewrite equation (58) as

$$0 \leq Q \sum_{x=0}^{x_i-1} \widetilde{T}(x_{i+1}, x) \leq u_{i+1} - s_{i+1} \pmod{Q} < Q \sum_{x=0}^{x_i} \widetilde{T}(x_{i+1}, x) \tag{60}$$

which, from equation (55), implies that $x_i^\dagger = x_i$. In equation (56), we can now replace $x_i^\dagger$ with $x_i$ and replace $u_{i+1}$ with the right side of equation (3), leading to the conclusion that $u_i^\dagger = u_i$.

# Appendix B: Proof that the map for a simulation of a discrete distribution is one-to-one and volume preserving

We can show that the map from $(x_i, y_i^*, u_i^*)$ to $(x_{i+1}, y_{i+1}^*, u_{i+1}^*)$ given by equations (22) to (24) is one-to-one by seeing that the following map is its inverse:

$$x_i^\dagger = \max\left\{x' : \sum_{x=0}^{x'-1} \widetilde{T}(x_{i+1}, x) \leq u_{i+1}^* - s_i^* \pmod{1}\right\} \tag{61}$$

$$y_i^{*\dagger} = \pi(x_i^\dagger) \left((u_{i+1}^* - s_i^* \pmod{1}) - \sum_{x=0}^{x_i^\dagger - 1} \widetilde{T}(x_{i+1}, x)\right) \Big/ \widetilde{T}(x_{i+1}, x_i^\dagger) \tag{62}$$

$$u_i^{*\dagger} = \sum_{x=0}^{x_{i+1}-1} T(x_i^\dagger, x) + T(x_i^\dagger, x_{i+1}) \frac{y_{i+1}^*}{\pi(x_{i+1})} \pmod{1} \tag{63}$$

That is, we will show that $x_i^\dagger = x_i$, $y_i^{*\dagger} = y_i$, and $u_i^{*\dagger} = u_i^*$. From equation (24) and the fact that $y_i^*$ is in $[0, \pi(x_i))$ we see that

$$u_{i+1}^* - s_i^* \pmod{1} = \sum_{x=0}^{x_i-1} \widetilde{T}(x_{i+1}, x) + \widetilde{T}(x_{i+1}, x_i) \frac{y_i^*}{\pi(x_i)} \tag{64}$$

since the right side will be in the interval $[0, 1)$. Substituting this in equation (61), we get

$$x_i^\dagger = \max\left\{ x' : \sum_{x=0}^{x'-1} \widetilde{T}(x_{i+1}, x) \leq \sum_{x=0}^{x_i-1} \widetilde{T}(x_{i+1}, x) + \widetilde{T}(x_{i+1}, x_i) \frac{y_i^*}{\pi(x_i)} \right\} = x_i \tag{65}$$

After replacing $x_i^\dagger$ with $x_i$ in equations (62) and (63), and using equations (64) and (23), we find that $y_i^{*\dagger} = y_i$ and $u_i^{*\dagger} = u_i^*$.

To see that volume is preserved by the map defined by equations (22) to (24) we can look at the Jacobian matrix for the continuous part of this map, from $(y_i^*, u_i^*)$ to $(y_{i+1}^*, u_{i+1}^*)$, over a region where $x_{i+1}$ is constant:

$$\begin{bmatrix} \dfrac{\partial y_{i+1}^*}{\partial y_i^*} & \dfrac{\partial u_{i+1}^*}{\partial y_i^*} \\ \dfrac{\partial y_{i+1}^*}{\partial u_i^*} & \dfrac{\partial u_{i+1}^*}{\partial u_i^*} \end{bmatrix} = \begin{bmatrix} 0 & \dfrac{\widetilde{T}(x_{i+1}, x_i)}{\pi(x_i)} \\ \dfrac{\pi(x_{i+1})}{T(x_i, x_{i+1})} & 0 \end{bmatrix} = \begin{bmatrix} 0 & \dfrac{T(x_i, x_{i+1})}{\pi(x_{i+1})} \\ \dfrac{\pi(x_{i+1})}{T(x_i, x_{i+1})} & 0 \end{bmatrix} \tag{66}$$

The absolute value of the determinant of this Jacobian matrix is one, so the map preserves volume.

## Appendix C: Proof that the map for a Metropolis-Hastings simulation of a discrete distribution is one-to-one and volume preserving

The map from $(x_i, y_i^*, u_i^*)$ to $(x_{i+1}, y_{i+1}^*, u_{i+1}^*)$ defined by equations (28) to (32) is its own inverse when $s_i^* = 0$, and in general its inverse is given by

$$\hat{x}_i^\dagger = \max\left\{ x' : \sum_{x=0}^{x'-1} S(x_{i+1}, x) \leq u_{i+1}^* - s_i^* \pmod{1} \right\} \tag{67}$$

$$a_i^\dagger = \left( (u_{i+1}^* - s_i^* \pmod{1}) - \sum_{x=0}^{\hat{x}_i^\dagger - 1} S(x_{i+1}, x) \right) \Big/ S(x_{i+1}, \hat{x}_i^\dagger) \tag{68}$$

$$x_i^\dagger = \begin{cases} \hat{x}_i^\dagger & \text{if } a_i^\dagger < a(x_{i+1}, \hat{x}_i^\dagger) \\ x_{i+1} & \text{otherwise} \end{cases} \tag{69}$$

$$y_i^{*\dagger} = \begin{cases} \pi(x_i^\dagger) \, a_i^\dagger / a(x_{i+1}, \hat{x}_i^\dagger) & \text{if } a_i^\dagger < a(x_{i+1}, \hat{x}_i^\dagger) \\ y_{i+1}^* & \text{otherwise} \end{cases} \tag{70}$$

$$u_i^{*\dagger} = \begin{cases} \displaystyle\sum_{x=0}^{x_{i+1}-1} S(x_i^\dagger, x) + S(x_i^\dagger, x_{i+1}) \, a(x_i^\dagger, x_{i+1}) \frac{y_{i+1}^*}{\pi(x_{i+1})} \pmod{1} & \text{if } a_i^\dagger < a(x_{i+1}, \hat{x}_i^\dagger) \\ u_{i+1}^* - s_i^* \pmod{1} & \text{otherwise} \end{cases} \tag{71}$$

We can show that $x_i^\dagger = x_i$, $y_i^{*\dagger} = y_i^*$, and $u_i^{*\dagger} = u_i^*$, separately for rejected and accepted proposals. I will do this here assuming $s_i^* = 0$; the general case follows from this, since adding $s_i^*$ to $u_i^*$ is itself one-to-one and volume-preserving.

If the proposal $\hat{x}_{i+1}$ is rejected — that is, if $a_{i+1} \geq a(x_i, \hat{x}_{i+1})$ — then $x_{i+1}^* = x_i^*$, $y_{i+1}^* = y_i^*$, and $u_{i+1}^* = u_i^*$, and it is easy to see that the inverse map above will also leave $x^*$, $y^*$, and $u^*$ unchanged.

If the proposal $\hat{x}_{i+1}$ is accepted, when $a_{i+1} < a(x_i, \hat{x}_{i+1})$, one can see that the way that $u_{i+1}^*$ is set ensures that $\hat{x}_i^\dagger = x_i$ — that is, the proposal will be the the previous state. This proposal will be accepted, since

$$a_i^\dagger = \left(u_{i+1}^* - \sum_{x=0}^{\hat{x}_i^\dagger - 1} S(x_{i+1}, x)\right) \Big/ S(x_{i+1}, \hat{x}_i^\dagger) = \left(u_{i+1}^* - \sum_{x=0}^{x_i - 1} S(x_{i+1}, x)\right) \Big/ S(x_{i+1}, x_i) = a(x_{i+1}, x_i) \frac{y_i^*}{\pi(x_i)}$$

$$< a(x_{i+1}, x_i) = a(x_{i+1}, \hat{x}_i^\dagger) \tag{72}$$

Since the proposal is accepted, $x_i^\dagger = x_i$. From $a_i^\dagger = a(x_{i+1}, x_i)\, y_i^* / \pi(x_i)$, one can also see that $y_i^{*\dagger} = y_i^*$. From equation (30), we see that

$$u_i^* = \sum_{x=0}^{\hat{x}_{i+1} - 1} S(x_i, x) + a_{i+1} S(x_i, x_{i+1}) \tag{73}$$

and from equation (31) we see that

$$a_{i+1} = a(x_i, x_{i+1})\, y_{i+1}^* / \pi(x_{i+1}) \tag{74}$$

from which it follows that $u_i^{*\dagger} = u_i^*$.

To confirm that the map defined by equations (28) to (32) preserves volume, we look at the Jacobian matrix for the continuous part of the map, from $(y_i^*, u_i^*)$ to $(y_{i+1}^*, u_{i+1}^*)$. For a rejected proposal, when $a_i \geq a(x_i, \hat{x}_i)$, this is the identity matrix. For an accepted proposal, the Jacobian matrix is as follows (noting that $\hat{x}_i = x_{i+1}$):

$$\begin{bmatrix} \dfrac{\partial y_{i+1}^*}{\partial y_i^*} & \dfrac{\partial u_{i+1}^*}{\partial y_i^*} \\ \dfrac{\partial y_{i+1}^*}{\partial u_i^*} & \dfrac{\partial u_{i+1}^*}{\partial u_i^*} \end{bmatrix} = \begin{bmatrix} 0 & \dfrac{\pi(x_{i+1})}{S(x_i, x_{i+1})\, a(x_i, x_{i+1})} \\ \dfrac{S(x_{i+1}, x_i)\, a(x_{i+1}, x_i)}{\pi(x_i)} & 0 \end{bmatrix} \tag{75}$$

$$= \begin{bmatrix} 0 & \dfrac{1}{\min\left[\dfrac{S(x_i, x_{i+1})}{\pi(x_{i+1})},\, \dfrac{S(x_{i+1}, x_i)}{\pi(x_i)}\right]} \\ \min\left[\dfrac{S(x_{i+1}, x_i)}{\pi(x_i)},\, \dfrac{S(x_i, x_{i+1})}{\pi(x_{i+1})}\right] & 0 \end{bmatrix} \tag{76}$$

Since the absolute value of the determinant of this Jacobian matrix is one, the map must preserve volume.

# Appendix D: Proof that the map for a simulation of a continuous distribution on an interval $(a, b)$ is one-to-one and volume preserving

The map from $(x_i^*, u_i^*, y_i^*, v_i^*)$ to $(x_{i+1}^*, u_{i+1}^*, y_{i+1}^*, v_{i+1}^*)$ defined by equations (38) to (41) is one-to-one because it has the following inverse:

$$x_i^{*\dagger} = \widetilde{F}_*^{-1}(x_{i+1}^*, u_{i+1}^* - s_i^* \ (\mathrm{mod}\ 1)) \tag{77}$$

$$u_i^{*\dagger} = F_*(x_i^{*\dagger}, x_{i+1}^*) \tag{78}$$

$$y_i^{*\dagger} = \pi^*(x_i^{*\dagger})(v_{i+1}^* - t_i^* \ (\mathrm{mod}\ 1)) \tag{79}$$

$$v_i^{*\dagger} = y_{i+1}^* / \pi^*(x_{i+1}^*) \tag{80}$$

Showing the $x_i^{*\dagger} = x_i^*$, $u_i^{*\dagger} = u_i^*$, $y_i^{*\dagger} = y_i^*$, and $v_i^{*\dagger} = v_i^*$ is straightforward except for issues involving $\pi^*(x^*)$ or $\widetilde{T}^*(x^*, x^{*\dagger})$ being zero. Division by zero in equation (80) will not occur because the state space excludes points where $\pi^*(x^*) = 0$, as no $y^*$ value is allowable with such an $x^*$. When $\widetilde{T}^*(x^*, x^{*\dagger})$ may be zero, we define $\widetilde{F}_*^{-1}(x^*, u^*)$ to be the largest $x^{*\prime}$ for which $\widetilde{F}_*(x^*, x^{*\prime}) = u^*$. $\widetilde{F}_*^{-1}$ will fail to invert $\widetilde{F}_*$ only where the reverse transition density is zero, but this cannot happen for a transition that was taken, for which the forward transition density must have been non-zero.

To see that the map defined by equations (38) to (41) preserves volume, we look at the determinant of its Jacobian matrix, which is

$$\begin{bmatrix} \frac{\partial x_{i+1}^*}{\partial x_i^*} & \frac{\partial u_{i+1}^*}{\partial x_i^*} & \frac{\partial y_{i+1}^*}{\partial x_i^*} & \frac{\partial v_{i+1}^*}{\partial x_i^*} \\ \frac{\partial x_{i+1}^*}{\partial u_i^*} & \frac{\partial u_{i+1}^*}{\partial u_i^*} & \frac{\partial y_{i+1}^*}{\partial u_i^*} & \frac{\partial v_{i+1}^*}{\partial u_i^*} \\ \frac{\partial x_{i+1}^*}{\partial y_i^*} & \frac{\partial u_{i+1}^*}{\partial y_i^*} & \frac{\partial y_{i+1}^*}{\partial y_i^*} & \frac{\partial v_{i+1}^*}{\partial y_i^*} \\ \frac{\partial x_{i+1}^*}{\partial v_i^*} & \frac{\partial u_{i+1}^*}{\partial v_i^*} & \frac{\partial y_{i+1}^*}{\partial v_i^*} & \frac{\partial v_{i+1}^*}{\partial v_i^*} \end{bmatrix} = \begin{bmatrix} \frac{\partial x_{i+1}^*}{\partial x_i^*} & \left[\widetilde{T}^*(x_{i+1}^*, x_i^*) + D\frac{\partial x_{i+1}^*}{\partial x_i^*}\right] & \times & \times \\ \frac{1}{T^*(x_i^*, x_{i+1}^*)} & \frac{D}{T^*(x_i^*, x_{i+1}^*)} & \times & \times \\ 0 & 0 & 0 & \frac{1}{\pi^*(x_i^*)} \\ 0 & 0 & \pi^*(x_{i+1}^*) & 0 \end{bmatrix} \tag{81}$$

Here, $D = (\partial/\partial x_{i+1}^*)\widetilde{F}_*(x_{i+1}^*, x_i^*)$ and $\times$ marks elements of the matrix that are irrelevant given the zero elements elsewhere. The determinant of this matrix has two non-zero terms, whose sum is

$$\left[\widetilde{T}^*(x_{i+1}^*, x_i^*) + D\frac{\partial x_{i+1}^*}{\partial x_i^*}\right] \frac{1}{T^*(x_i^*, x_{i+1}^*)} \frac{1}{\pi^*(x_i^*)} \pi^*(x_{i+1}^*) - \frac{\partial x_{i+1}^*}{\partial x_i^*} \frac{D}{T^*(x_i^*, x_{i+1}^*)} \frac{1}{\pi^*(x_i^*)} \pi^*(x_{i+1}^*)$$

$$= \widetilde{T}^*(x_{i+1}^*, x_i^*) \frac{1}{T^*(x_i^*, x_{i+1}^*)} \frac{1}{\pi^*(x_i^*)} \pi^*(x_{i+1}^*) = \frac{\pi^*(x_{i+1}^*)\widetilde{T}^*(x_{i+1}^*, x_i^*)}{\pi^*(x_i^*)T^*(x_i^*, x_{i+1}^*)} = 1 \tag{82}$$

Since the determinant is one, the map preserves volume.

# Appendix E: R Program for Ising Model Simulations

```
# SETUP FOR ISING MODEL.  Stores information in global variables for later use.

ising_setup <- function (n_rows, n_cols)
{ n_rows <<- n_rows; n_cols <<- n_cols; n_spins <<- n_rows * n_cols; n_neighbors <<- 4
  neighbor <<- matrix (NA, n_spins, n_neighbors)
  for (j in 1:n_spins)
  { neighbor[j,1] <<- if (j%%n_rows==0) j-(n_rows-1) else j+1
    neighbor[j,2] <<- if (j%%n_rows==1) j+(n_rows-1) else j-1
    neighbor[j,3] <<- if ((j-1)%/%n_rows==n_cols-1) j-n_rows*(n_cols-1) else j+n_rows
    neighbor[j,4] <<- if ((j-1)%/%n_rows==0) j+n_rows*(n_cols-1) else j-n_rows
  }
}

# VECTORIZED MCMC FOR ISING MODEL.  Method is "s" for standard simulation with multiple
# random streams, "c" for coupled simulation with a single stream, and "p" for permuation
# MCMC with a single stream.  Initial values for states, u, and yf (y divided by pi) can be
# specified, or left to be generated randomly.  The "random" numbers can be supplied via the
# s argument (of dimensions n_iters by n_spins), and are otherwise set here using runif.
# If the rev argument is TRUE, the reverse map is applied.  The value returned is an array
# of states at all iterations, with attributes giving end values of states, u, and yf.

vec_ising <- function (beta, method, n_chains, n_iters, states=NULL, u=NULL, yf=NULL,
                       s=NULL, rev=FALSE)
{ if (is.null(states))
    states <- matrix (sample (c(-1,1), n_chains*n_spins, replace=TRUE), n_spins, n_chains)
  if (is.null(u) && method=="p") u <- runif (n_chains)
  if (is.null(yf) && method=="p") yf <- runif (n_chains)
  if (is.null(s) && (method=="p" || method=="c"))
    s <- matrix(runif(n_iters*n_spins),n_iters,n_spins)

  if (rev) s <- s[n_iters:1,,drop=FALSE]

  results <- array(dim=c(n_spins,n_chains,n_iters))

  for (i in 1:n_iters)
  { for (j in { if (rev) n_spins:1 else 1:n_spins })
    {
      if (method=="p" && rev) u <- (u - s[i,j]) %% 1    # Take s from u first for reverse map

      neighbor_sum <- rep(0,n_chains)
      for (k in 1:n_neighbors) neighbor_sum <- neighbor_sum + states[neighbor[j,k],]

      if (method=="s") u <- runif(n_chains)
      if (method=="c") u <- s[i,j]
      if (method=="p") old_spin01 <- states[j,] == 1   # Old values of spins in 0/1 form

      p0 <- 1 / (1 + exp(2*beta*neighbor_sum))          # Probabilities of setting spin to 0
      spin01 <- p0 <= u                                 # New values of spins in 0/1 form
      states[j,] <- 2*spin01 - 1                        # New values of spins in -1/+1 form

      if (method=="p")
      { t <- spin01*(1-p0) + (1-spin01)*p0              # Probabilities of new spin values
```

```r
          old_t <- old_spin01*(1-p0) + (1-old_spin01)*p0  # Probabilities of old spin values
          old_yf <- yf
          yf <- (u - p0*spin01) / t                       # New yf (equal to y/pi) from eq (23)
          u <- p0*old_spin01 + old_t*old_yf               # New u (except for + s) from eq (24)
          if (!rev) u <- (u + s[i,j]) %% 1                # Add s to u last for non-reverse map
        }
    }
    results[,,i] <- states
  }

  structure (results, end_states=states, end_u=u, end_yf=yf)
}

# COMPUTE THE ENERGY OF A SPIN CONFIGURATION.

ising_energy <- function (spins)
{ energy <- 0
  for (j in 1:n_spins) energy <- energy - spins[j] * sum(spins[neighbor[j,]])
  energy / 2  # Divide by two because each term in the energy was added twice above
}

# PLOT THE ENERGY ALONG ALL CHAINS.

energy_plot <- function (res)
{ n_chains <- dim(res)[2]; n_iters <- dim(res)[3]

  E <- matrix(NA,n_iters,n_chains)
  for (i in 1:n_iters) for (k in 1:n_chains) E[i,k] <- ising_energy(res[,k,i])
  E_bar <- colMeans(E)

  cat ("Energy: mean ",round(mean(E_bar),3), " +- ",round(sd(E_bar)/sqrt(n_chains),3),
       "\n",sep="")

  plot (c(1,n_iters), c(-2*n_spins, 2*n_spins), type="n", xlab="Iteration", ylab="")
  for (k in 1:n_chains) points (E[,k], col=k, pch=20)
}

# PLOT THE MAGNETIZATION (SUM OF SPINS) ALONG ALL CHAINS.

magnetization_plot <- function (res)
{ n_chains <- dim(res)[2]; n_iters <- dim(res)[3]

  M <- matrix(NA,n_iters,n_chains)
  for (i in 1:n_iters) for (k in 1:n_chains) M[i,k] <- sum(res[,k,i])
  M_bar <- colMeans(M); M_abs_bar <- colMeans(abs(M))

  cat ("Magnetization, mean ",round(mean(M_bar),3), " +- ",round(sd(M_bar)/sqrt(n_chains),3),
       ", mean abs ",round(mean(M_abs_bar),3), " +- ",round(sd(M_abs_bar)/sqrt(n_chains),3),
       "\n",sep="")

  plot (c(1,n_iters), c(-n_spins, n_spins), type="n", xlab="Iteration", ylab="")
  for (k in 1:n_chains) points (apply(res[,k,],2,sum), col=k, pch=20)
}
```

# Appendix F: R Program for Truncated Normal Simulations

```
# VECTORIZED GIBSS SAMPLING FOR TRUNCATED NORMAL MODEL.  Method is "s" for standard simulation
# "c" for coupled simulation, and "p" for permuation MCMC.  Initial values and the random
# stream can be specified, or left to be generated randomly.  If the rev argument is TRUE,
# the map is reversed.  The value returned is an array of states at all iterations, with
# attributes giving end values of states and u.  Uses global rho, lower, and upper variables.

vec_trunc_normal_gibbs <- function (method, n_chains, n_iters, states=NULL, u=NULL,
                                    s=NULL, rev=FALSE)
{ if (is.null(states))
  { states <- matrix (NA, 2, n_chains)
    for (j in 1:2) states[j,] <- runif(n_chains, lower[j], upper[j])
  }
  if (is.null(u) && method=="p") u <- runif (n_chains)
  if (is.null(s) && (method=="p" || method=="c")) s <- matrix(runif(2*n_iters),n_iters,2)
  if (rev) s <- s[n_iters:1,,drop=FALSE]

  results <- array(dim=c(2,n_chains,n_iters))

  for (i in 1:n_iters)
  { for (j in { if (rev) 2:1 else 1:2 })
    { if (method=="p" && rev) u <- (u - s[i,j]) %% 1
      if (method=="s") u <- runif(n_chains)
      if (method=="c") u <- s[i,j]
      if (method=="p") old <- states[j,]
      cond_sd <- sqrt(1-rho^2); cond_mean <- rho * states[3-j,]
      ql <- pnorm (lower[j], cond_mean, cond_sd); qu <- pnorm (upper[j], cond_mean, cond_sd)
      states[j,] <- # max & min below not usually needed - just in case of numerical issue
        pmax (lower[j], pmin (upper[j], qnorm (ql + (qu-ql)*u, cond_mean, cond_sd)))
      if (method=="p")
      { u <- (pnorm(old,cond_mean,cond_sd) - ql) / (qu-ql)
        if (!rev) u <- (u + s[i,j]) %% 1
      }
    }
    results[,,i] <- states
  }
  structure (results, end_states=states, end_u=u)
}

# VECTORIZED METROPOLIS SAMPLING FOR TRUNCATED NORMAL MODEL.  Arguments are as above, but
# with no "rev" argument, an extra optional "yf" argument, and an extra optional "delta"
# argument that specifies random-walk proposal offsets for each variable update.

vec_trunc_normal_met <- function (method, n_chains, n_iters, states=NULL, u=NULL, yf=NULL,
                                  s=NULL, delta=NULL)
{ if (is.null(states))
  { states <- matrix (NA, 2, n_chains)
    for (j in 1:2) states[j,] <- runif(n_chains, lower[j], upper[j])
  }
  if (is.null(u) && method=="p") u <- runif (n_chains)
  if (is.null(yf) && method=="p") yf <- runif(n_chains)
  if (is.null(s) && (method=="p" || method=="c")) s <- matrix(runif(2*n_iters),n_iters,2)
  if (is.null(delta)) delta <- matrix(runif(2*n_iters,0,1),n_iters,2)
```

```
  results <- array(dim=c(2,n_chains,n_iters))

  for (i in 1:n_iters)
  { for (j in 1:2)
    { if (method=="s") u <- runif(n_chains)
      if (method=="c") u <- s[i,j]
      old <- states[j,]
      cond_sd <- sqrt(1-rho^2); cond_mean <- rho * states[3-j,]
      proposal <- (states[j,]+delta[i,j]) * (u<1/2) + (states[j,]-delta[i,j]) * (u>=1/2)
      a <- (2*u) %% 1
      ar <- dnorm(proposal,cond_mean,cond_sd) / dnorm(old,cond_mean,cond_sd)
      ap <- (proposal >= lower[j]) * (proposal <= upper[j]) * pmin(1,ar)
      accept <- (a < ap)
      states[j,] <- accept*proposal + (1-accept)*old
      if (method=="p")
      { u[accept] <- ((u<1/2)/2 + pmin(1,1/ar)*yf/2)[accept]; yf[accept] <- (a/ap)[accept]
        u <- (u + s[i,j]) %% 1
      }
    }
    results[,,i] <- states
  }
  results
}

# PLOT X1 VERSUS X2 FOR SIMULATIONS.

bi_plot <- function (res, from=11, xlim=c(lower[1],upper[1]), ylim=c(lower[2],upper[2]))
{ n_chains <- dim(res)[2]; n_iters <= dim(res)[3]
  plot (c(), type="n", xlab="x1", ylab="x2", xlim=xlim, ylim=ylim)
  for (k in 1:n_chains) points (res[1,k,from:n_iters], res[2,k,from:n_iters], pch=20, col=k)
}

# PLOT X1 OR X2 VERSUS ITERATION NUMBER FOR SIMULATIONS.

uni_plot <- function (res, j, ylim=c(lower[j],upper[j]))
{ n_chains <- dim(res)[2]; n_iters <- dim(res)[3]
  plot (c(),type="n",xlab="Iteration",ylab=paste("x",j,sep=""),xlim=c(1,n_iters),ylim=ylim)
  for (k in 1:n_chains) points (res[j,k,], pch=20, col=k)
}

# FIND MEAN ESTIMATES AND STANDARD ERRORS FROM SIMULATIONS.

estimates <- function (res, from=11)
{ n_chains <- dim(res)[2]; n_iters <- dim(res)[3]
  ave <- ave_sq <- matrix(NA,2,n_chains)
  for (k in 1:n_chains)
  { ave[,k] <- rowMeans(res[,k,from:n_iters]);
    ave_sq[,k] <- rowMeans(res[,k,from:n_iters]^2)
  }
  round (cbind (ave=apply(ave,1,mean),       se=apply(ave,1,sd)/sqrt(n_chains),
                ave_sq=apply(ave_sq,1,mean), se_sq=apply(ave_sq,1,sd)/sqrt(n_chains)), 4)
}
```

# Appendix G: R Program for Importance Sampling Test Simulations

```
# LOG PROBABILITY DENSITY FUNCTION FOR TEST DISTRIBUTION.

istest_log_density <- function (x) dnorm(x[1],0,1,log=TRUE) + dnorm(x[2],x[1]^2-1,1,log=TRUE)

# VECTORIZED PERMUTATION METROPOLIS FOR TEST DISTRIBUTION.  Does random-walk Metropolis updates
# for both variables at once.  If delta is not supplied, it's normal with given proposal_sd.

vec_istest_met <- function (n_chains, n_iters, states=NULL, u=NULL, yf=NULL,
                            s=NULL, delta=NULL, rev=FALSE, proposal_sd)
{
  if (is.null(states)) states <- matrix (rnorm(n_chains*2,0,1), 2, n_chains)
  if (is.null(u))      u <- runif(n_chains)
  if (is.null(yf))     yf <- runif(n_chains)
  if (is.null(s))      s <- runif(n_iters)
  if (is.null(delta))  delta <- matrix(rnorm(n_iters*2,0,proposal_sd),n_iters,2)
  if (rev) { s <- s[n_iters:1]; delta <- delta[n_iters:1,,drop=FALSE] }

  results <- array(dim=c(2,n_chains,n_iters))

  for (i in 1:n_iters)
  {
    if (rev) u <- (u - s[i]) %% 1

    proposal <- t (t(states+delta[i,]) * (u<1/2) + t(states-delta[i,]) * (u>=1/2))
    a <- (2*u) %% 1
    ar <- exp (apply(proposal,2,istest_log_density) - apply(states,2,istest_log_density))
    ap <- pmin(1,ar)
    accept <- (a < ap)
    states <- t (accept*t(proposal) + (1-accept)*t(states))
    u[accept] <- ((u<1/2)/2 + pmin(1,1/ar)*yf/2)[accept]
    yf[accept] <- (a/ap)[accept]

    if (!rev) u <- (u + s[i]) %% 1

    results[,,i] <- states
  }

  structure (results, end_states=states, end_u=u, end_yf=yf)
}

# METROPOLIS IMPORTANCE SAMPLING FOR TEST DISTRIBUTION.  N is number of points, M is maximum
# number of transitons, mu and sd are for normal start state distribution, and proposal_sd
# is for the normal random-walk Metropolis proposals.

istest_met_is <- function (N, M, mu, sd, proposal_sd)
{
  pr <- rep(NA,M+1); w <- rep(NA,N); beta <- matrix (NA,N,2)
  states <- matrix(NA,2,M+1); s <- runif(M); delta <- matrix (rnorm(M*2,0,proposal_sd), M, 2)

  for (i in 1:N)
  {
    k <- if (M==0) 0 else sample(0:M,1)  # Pick number of transitions from start to end state
```

```r
    states[,k+1] <- rnorm(2,mu,sd)       # Random start state
    u <- runif(1); yf <- runif(1)        # Random start values for u and yf

    if (k<M)                             # Do a forward simulation from the start state on
      states[,(k+2):(M+1)] <-
        vec_istest_met (proposal_sd, 1, M-k, cbind(states[,k+1]), u, yf,
                     s=s[(k+1):M], delta=delta[(k+1):M,,drop=FALSE], rev=FALSE) [,1,]

    if (k>0)                             # Do a reverse simulation before the start state
      states[,1:k] <-
        vec_istest_met (proposal_sd, 1, k, cbind(states[,k+1]), u, yf,
                     s=s[1:k], delta=delta[1:k,,drop=FALSE], rev=TRUE) [,1,]

    beta[i,] <- states[,M+1]             # Importance sampling point is the end state

    for (k in 0:M)
      pr[k+1] <- sum(dnorm(states[,k+1],mu,sd,log=TRUE))      # Compute log probabilities
                 - istest_log_density(states[,k+1])           #   of possible start states
    w[i] <- - (max(pr) + log(sum(exp(pr-max(pr)))) - log(M+1)) # Importance weight: minus the
  }                                                            #   log of average probability

  list (w=w, beta=beta)
}

# ANALYSE IMPORTANCE SAMPLING OUTPUT.  Takes as arguments a vector of log weights and a
# matrix of beta values.  Prints various estimates with standard errors.

is_analysis <- function (w, beta)
{
  N <- length(w)
  mx <- max(w)
  w <- exp (w-mx)

  cat ("\nEstimated log normalizing constant:", round(mx+log(mean(w)),3),
       "+-", round(sd(w)/sqrt(N)/mean(w),3), "\n\n")

  w <- w / mean(w)

  z <- sum(w)
  ave <- colSums(w*beta) / z
  ave_sq <- colSums(w*beta^2) / z

  cat ("Adjusted sample size:", N, "/", 1+var(w), "=", N/(1+var(w)), "\n\n")

  beta_offset <- t (t(beta) - ave)
  ave_se <- sqrt (colSums((w*beta_offset)^2) / N^2)
  beta_sq_offset <- t (t(beta^2) - ave_sq)
  ave_sq_se <- sqrt (colSums((w*beta_sq_offset)^2) / N^2)

  print(cbind (ave=ave, se=ave_se, ave_sq=ave_sq, sq_se=ave_sq_se))
}
```